\documentclass[10pt,conference]{IEEEtran}%
\usepackage{graphicx}
\usepackage{algorithm}
\usepackage{algorithmic}
\usepackage{amsmath}
\usepackage{amsfonts}
\usepackage{amssymb}
\usepackage{psfrag}
\usepackage{cite}
\usepackage{mycolor}%
\providecommand{\U}[1]{\protect\rule{.1in}{.1in}}
\IEEEoverridecommandlockouts
\newtheorem{theorem}{Theorem}
\newtheorem{lemma}{Lemma}

\newtheorem{definition}{Definition}

\newenvironment{allauthors}{
\title{\Huge A Fundamental Characterization of Stability in Broadcast Queueing Systems}
\author{Chan Zhou$^*$ and Gerhard Wunder$^*$\thanks{$^*$Fraunhofer German-Sino Lab for Mobile Communications,
Heinrich-Hertz-Institut, Einstein-Ufer 37, D-10587 Berlin,
Germany, \textbf{Email}: \{zhou,wunder\}@hhi.fhg.de. Parts of this
work was supported by the German Ministry for Education and
Research under Grant FK 01 BU 566 (the \emph{ScaleNet} project).
Parts were presented at the Forty-Sixth Annual Allerton Conference
on Communication, Control, and Computing in  Monticello (USA,
2008) and the IEEE International Symposium on Information Theory
(ISIT) in Seoul (Korea, 2009).}} \maketitle
\begin{abstract}
Stability with respect to a given scheduling policy has become an
important issue for wireless communication systems, but it is hard
to prove in particular scenarios. In this paper two simple
conditions for stability in broadcast channels are derived, which
are easy to check. Heuristically, the conditions imply that if the
queue length in the system becomes large, the rate allocation is
always the solution of a weighted sum rate maximization problem.
Furthermore, the change of the weight factors between two time
slots becomes smaller and the weight factors of the users, whose
queues are bounded while the other queues expand, tend to zero.
Then it is shown that for any mean arrival rate vector inside the
ergodic achievable rate region the system is stable in the strong
sense when the given scheduling policy complies with the
conditions. In this case the policy is so-called
throughput-optimal. Subsequently, some results on the necessity of
the presented conditions are provided. Finally, in several
application examples it is shown that the results in the paper
provide a convenient way to verify the throughput-optimal
policies.
\end{abstract}}
\\
\begin{document}
\allauthors

\section{Introduction}

\label{section:introduction}

In wireless communication systems input data packets arrive randomly at the
transmitter and queue up in a buffer awaiting the transmission. The
transmission rate for each user is determined by a scheduling policy which
acts according to the system state. One major challenge for the policy design
is to improve the system throughput by taking advantage of the channel
variations. However, policies which consider only the channel state will cause
unfairness among the users due to the randomness in the data traffic. Even
more severely, some users may suffer from long queueing delay or even buffer overflow.

In order to avoid this situation, the scheduling policy can take the queue
states into account. An essential requirement then is that the queues are kept
finite for all users and do not blow up over time, such that the queueing
system is stable. The stability of the queueing system is determined by the
arrival traffic, the transmission capacity and the applied scheduling policy.
A scheduling policy is called \emph{throughput-optimal} if it keeps the system
stable for any set of mean arrival rates that lies in the ergodic achievable
rate region. Throughput-optimality is a desirable feature of scheduling
policies, since the system can offer a maximal possible traffic load and keeps
all queues stable at the same time.

There already exist a number of throughput-optimal scheduling policies in
previous work, e.g. the Maximum Weight Matching (MWM) policy
\cite{Tassiulas:1992, McKeown:1999,NeeModRoh:2003, YehCoh:2003}, the
exponential rule \cite{Shakkottai:2001}, the Queue Proportional Scheduling
(QPS) \cite{Seong:2006}, the Idle State Prediction Scheduling (ISPS)
\cite{EURASIP_ISP}. Throughput-optimality of these policies is proven using
techniques which are adapted to the particular policy and in fact the proofs
can be rather involved
\cite{Tassiulas:1992,Shakkottai:2001,Seong:2006,EURASIP_ISP}. Eryilmaz et al.
provided some sufficient conditions under which the scheduling policies are
throughput-optimal\cite{Eryilmaz:2005}. However, these conditions are quite
restricted and do not include all throughput-optimal scheduling policies (e.g.
exponential rule, QPS, ISPS).

In this paper, we consider scheduling policies in a broadcast system and give
general conditions for their throughput-optimality. The scheduling policy is
formulated as the solution of a weighted sum rate maximization problem
differing only in the choice of the weight factors. Then we show that
throughput-optimality can be verified solely by checking characteristics of
the weight factors. It is shown that the weight factors of a
throughput-optimal scheduling policy only need to satisfy two conditions:

\begin{enumerate}
\item If the total queue length in the system becomes large, the change of the
weight factors between two time slots tends to zero.

\item If the total queue length in the system becomes large, the weight
factors of "nonurgent" users, whose queues are bounded while the other queues
expand, tend to zero.
\end{enumerate}

The proofs are derived by using some special Lyapunov functions in connection
with theorems in differential geometry. It is proven that the presented
conditions are also necessary and indeed cover all queue-length based
throughput-optimal scheduling policies. We apply these results to well-known
scheduling policies and show their throughput-optimality.

The remainder of this paper is organized as follows: Section
\ref{section:system} describes the system model including some
assumptions on the rate and arrival processes. In this section we
also introduce the definition of stability and
throughput-optimality. General sufficient conditions for
throughput-optimality are presented in Section
\ref{section:stability}. The necessity of these conditions is
proven in Section \ref{section:necessity}. Some applications of
our results are shown in Section \ref{section:applications}.
Finally we conclude in Section \ref{section:conclusions}.

\textbf{Notations}: We use boldface letters to denote vectors and common
letters with subscript are the elements. $\Vert\mathbf{x}\Vert_{i}$ denotes
the $l_{i}$-norm of the vector $\mathbf{x}$ and $\Vert\mathbf{x}\Vert$ is a
arbitrary norm of $\mathbf{x}$. $\mathbb{E}\{x\}$ denotes the expected value
of random variable $x$. Furthermore we use $\mathcal{A}^{c}$ to denote the
complement of a set $\mathcal{A}$. The probability function is denoted as
$Pr\{\cdot\}$. The indicator $\mathbb{I\{\cdot\}}$ equals 1 if the argument is
true and equals 0 otherwise.

\section{System Model}

\label{section:system}

\subsection{Physical layer}

We consider a single cell downlink system in which a base station
simultaneously supplies $M$ mobile users. The channel between the
base station and each user is assumed to be constant within a time
slot and varies from one time slot to another in a stationary
i.i.d. manner. The channel state of user $i$ in the $n$-th time
slot is denoted as $h_{i}(n)\in\mathcal{S}$, where $\mathcal{S}$
is an arbitrary countable or uncountable set, and all channel
states of the user set $\mathcal{M}:=\{1,...,M\}$ are collected in
the vector $\mathbf{h}(n)\in\mathcal{S}^{M}$. Here the set
$\mathcal{S}$ is used to indicate that the general approach is not
restricted to a specific transmission scheme. For example in a
MIMO system the channel state can be described as a matrix of
complex channel gains such that $h_{i}(n)\in
\mathbb{C}^{n_{r}n_{t}}$ where $n_{r}$, $n_{t}$ are the number of
transmit and receive antennas at the base station and mobiles,
respectively. Likewise for an OFDM system the channel state can be
defined as a vector of complex channel gains on each subcarrier
$h_{i}(n)\in\mathbb{C}^{K}$ where $K$ is the number of
subcarriers. Furthermore we assume that the channel state
information is, either perfectly or partially, accessible at the
receiver and the transmitter. In the $n$-th time slot the data is
transmitted through the channel at the rate
$\mathbf{r}(n)\in\mathbb{R}_{+}^{M}$ with the transmission power
$\mathbf{p}(n)\in\mathbb{R}_{+}^{M}$.

We define a resource allocation policy $\mathcal{P}$ with rate allocation
$\mathbf{r}^{\mathcal{P}}(n)$ and power allocation $\mathbf{p}^{\mathcal{P}%
}(n)$. The allocated rate satisfies
$\mathbf{r}^{\mathcal{P}}(n)\in \mathcal{C}\left(
\mathbf{h}(n),\mathbf{p}^{\mathcal{P}}(n)\right)  $, where
$\mathcal{C}\left(  \mathbf{h}(n),\mathbf{p}(n)\right)  $ is the
achievable rate region of the broadcast system with the channel
state $\mathbf{h}(n)$ and the transmission power $\mathbf{p}(n)$.
We don't specify the achievable rate region $\mathcal{C}\left(
\mathbf{h}(n),\mathbf{p}(n)\right)  $ so that the results in the
paper can be applied to different transmission schemes.

The entire achievable rate region of a system with maximum power constraint
$\hat{P}$ in time slot $n$ is given by
\begin{equation}
\mathcal{C}\left(  \mathbf{h}(n),\hat{P}\right)  := \bigcup_{\mathcal{P}%
\in\Omega} \mathcal{C}\left(  \mathbf{h}(n),\mathbf{p}^{\mathcal{P}%
}(n)\right)  ,
\end{equation}
where $\Omega$ is the set of feasible policies
\[
\Omega:=\left\{  \tilde{\mathcal{P}}:\left\Vert \mathbf{p}^{\tilde
{\mathcal{P}}}\left(  n\right)  \right\Vert _{1}\leq\hat{P},\forall n\right\}
.
\]
In general, the rate region $\mathcal{C}\left(
\mathbf{h}(n),\hat{P}\right) $ might be non-convex; then we
consider the convex hull of the achievable rate region. Any point
on the convex hull of the region $\mathcal{C}(
\mathbf{h}(n),\hat{P}) $ is a solution of the maximization problem
\begin{equation}
\mathbf{r}(\boldsymbol{\mu,}\mathbf{h}(n))=\underset{\tilde{\mathbf{r}}%
\in\mathcal{C}\left(  \mathbf{h}(n),\hat{P}\right)  }{\arg\max}\boldsymbol{\mu
}^{T}\tilde{\mathbf{r}}, \label{eqn:weighted_sum_rate_ins}%
\end{equation}
where $\boldsymbol{\mu}\in\mathbb{R}_{+}^{M}$ is the set of weight
factors. It is important to stress that even though the weight
factors are fixed and independent of the channel state
$\mathbf{h}(n)$, the rate allocation
$\mathbf{r}(\boldsymbol{\mu,}\mathbf{h}(n))$ depends on the
channel state due to the maximization problem over $\mathcal{C}(
\mathbf{h}(n),\hat{P})$ in (\ref{eqn:weighted_sum_rate_ins}).
Observe that weight factors $\boldsymbol{\mu }$ also represent a
normal vector of a supporting hyperplane which is tangential to
the convex hull at the point $\mathbf{r}(\boldsymbol{\mu},
\mathbf{h}(n))$.

Then, the ergodic achievable rate region is defined as
\begin{align}
&  \mathcal{C}(\hat{P})\nonumber\\
:=  &  \bigcap_{\left\Vert \boldsymbol{\mu}\right\Vert =1}\left\{  {r}%
_{1},...,{r}_{M}:\boldsymbol{\mu}^{T}{\mathbf{r}}\leq\mathbb{E}\left\{
\underset{\tilde{\mathbf{r}}\in\mathcal{C}\left(  \mathbf{h}(n),\hat
{P}\right)  }{\max}\boldsymbol{\mu}^{T} \tilde{\mathbf{r}} \right\}  \right\}
. \label{eqn:C_erg}%
\end{align}
Note that the definition (\ref{eqn:C_erg}) coincides with the
capacity region given in
\cite{Li_Goldsmith_2001,Viswanathan_Downlink:2003}, when
$\mathcal{C}\left( \mathbf{h}(n),\mathbf{p}(n)\right)  $ is the
capacity region of parallel degraded channels. Furthermore, it is
shown in \cite{NeeModRoh:2003, YehCoh:2003, Wireless_Exponential}
that the region is also the maximal achievable stability region.
In general the
characterization of the achievable rate region $\mathcal{C}(\mathbf{h}(n),%
\hat{P})$ and the solution of the weighted sum rate maximization
problem is complicated. Considering practical constraints such as
finite code and modulation scheme and imperfect channel state
information, the ergodic achievable rate region $\mathcal{C}(
\hat{P}) $ can be much smaller than the information-theoretical
capacity region. Thereby the instantaneous achievable rate region
$\mathcal{C}( \mathbf{h}(n),\hat{P}) $ might be non-convex or even
be restricted to a set of some discrete rate points. The results
in this paper can also be applied to these practical systems as
long as a solution of the problem in
(\ref{eqn:weighted_sum_rate_ins}) exists (i.e. for an discrete
OFDMA system described in \cite{EURASIP_2007}).

\begin{figure}[h]
\begin{center}
\psfrag{boundary of Cxxxxxxxx}{\tiny boundary of $\mathcal{C}(
\hat{P}) $} \psfrag{re}{\footnotesize
$\mathbf{r}_E(\boldsymbol{\mu})$} \psfrag{u}{\footnotesize$\boldsymbol{\mu}$}
\includegraphics[width=1\linewidth]{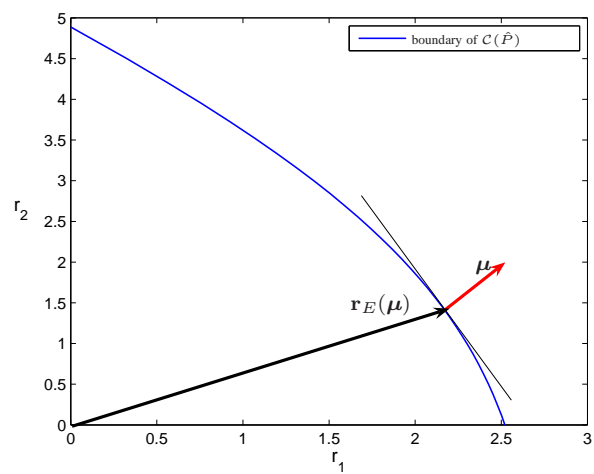}
\end{center}
\caption{The solution of the weight maximization problem
(\ref{eqn:weighted_sum_rate_erg}) is a point on the boundary of the ergodic
achievable rate region $\mathcal{C}( \hat{P}) $. The vector of weight factors
$\boldsymbol{\mu}$ can be interpreted as the normal vector of the boundary at
the obtained point}%
\label{fig_rate_region}%
\end{figure}

It is easy to show that the ergodic achievable region given in
(\ref{eqn:C_erg}) is convex. According to the convexity, any boundary point of
the region $\mathcal{C}(\hat{P})$ is a solution of the problem
\begin{equation}
\mathbf{r}_{E}(\boldsymbol{\mu})=\underset{\tilde{\mathbf{r}}\in
\mathcal{C}\left(  \hat{P}\right)  }{\arg\max}\boldsymbol{\mu}^{T}%
\tilde{\mathbf{r}}. \label{eqn:weighted_sum_rate_erg}%
\end{equation}
Likewise here the weight factors $\boldsymbol{\mu}$ also represents the normal
vector of the boundary at the point $\mathbf{r}_{E}\left(  \boldsymbol{\mu
}\right)  $ (see Fig.\ref{fig_rate_region}).

\subsection{Medium Access Control (MAC) layer}

Assuming that the transmission is time-slotted, data packets arrive randomly
at the MAC and queue up in a buffer reserved for each user $i\in\mathcal{M}$.
Simultaneously the data is read out from the buffers according to the system
state, i.e., the random channel state and the current queue lengths. Thus, the
system can be modeled as a queueing system with random processes reflecting
the arrival and the departure of data packets.

Denoting the queue state of the $i$-th buffer in time slot $n\in\mathbb{N}$ by
$q_{i}\left(  n\right)  $ and arranging all queue states in the vector
$\mathbf{q}(n)\in\mathbb{R}_{+}^{M}$, the evolution of the queueing system can
be written as
\begin{equation}
\label{eqn:markov_chain}\mathbf{q}\left(  n+1\right)  =\left[  \mathbf{q}%
\left(  n\right)  -\mathbf{r}\left(  n\right)  +\mathbf{a}\left(  n\right)
\right]  ^{+},
\end{equation}
where $[{x}]_{i}^{+}=\max\{0,x_{i}\}$, $\forall i\in\mathcal{M}$. Vector
$\mathbf{a}\left(  n\right)  \in\mathbb{R}_{+}^{M}$ is a random vector
denoting the amount of arrival packets during the $n$-th time slot and vector
$\mathbf{r}\left(  n\right)  \in\mathbb{R}_{+}^{M}$ is the amount of
transmitted data.

Without loss of generality we set the length of a time slot $T=1$ so that
$\mathbf{a}\left(  n\right)  $ and $\mathbf{r}\left(  n\right)  $ are equal to
the arrival and transmission rate during the time slot $n$. We assume that the
size of a data packet is constant. To simplify the notation we set the packet
size to $1$ unit without loss of generality.

Eqn.(\ref{eqn:markov_chain}) can also be formulated as
\begin{equation}
\mathbf{q}\left(  n+1\right)  =\mathbf{q}\left(  n\right)  +\mathbf{a}\left(
n\right)  -\mathbf{r}\left(  n\right)  +\mathbf{z}\left(  n\right)  ,
\end{equation}
with
\begin{align*}
&  z_{i}(n) = \left\{
\begin{tabular}
[c]{l}%
$0$ \;\;\;\;\;\;\;\;\;\;\;\;\;\;\;\;\;\; $q_{i}(n)+a_{i}(n)-r_{i}(n)\geq0$\\
$r_{i}(n)-q_{i}(n)-a_{i}(n)$ \; otherwise
\end{tabular}
\ \right.
\end{align*}

We assume that the sequence of arrival bits forms an i.i.d. sequence of
variables over time. For technical reasons we assume that the arrival bits
$a_{i}(n)$ are uniformly bounded by some real constant $C_{a}>0$.

The transmission rate $\mathbf{r}\left(  n\right)  $ is determined by the
applied scheduling policy. We consider scheduling policies which are
independent of the time index and define the policies as the mapping from the
cartesian product of the set of channel gains $\mathbf{h}\left(  n\right)  $
and queue lengths $\mathbf{q}\left(  n\right)  $ to the set of transmission
rates. The rate allocated by policy $\mathcal{P}$ is denoted as $\mathbf{r}%
^{\mathcal{P}}(\mathbf{h}(n),\mathbf{q}(n))$. Further we make the technical
assumption that the maximum transmission rate $\mathbf{r}^{\mathcal{P}%
}(\mathbf{h}(n),\mathbf{q}(n))$ are uniformly bounded by some real constant
$C_{r}>0$. Under these assumptions, the considered queueing system can be
modeled as a $\psi$-irreducible Markov chain with $\psi$-irreducible measure
$\delta_{0}$ where $\delta_{0}$ denotes a Dirac measure at zero
\cite{Meyn:1993}.

\subsection{Stability}

The stability of an $\psi$-irreducible Markov chain can be defined in
different manners. We first introduce the definitions of \emph{recurrent} and
\emph{transient} Markov chain as given in \cite{Meyn:1993}. These definitions
are based on the measure of the occupation time
\[
\eta_{\mathcal{A}}:=\sum_{n=1}^{\infty}\mathbb{I}\left(  \mathbf{q}\left(
n\right)  \in\mathcal{A}\right)
\]
which gives the number of visits in a set $\mathcal{A} \in\mathbb{R}^{M}_{+}$
by a Markov chain after time zero.

\begin{definition}
A Markov chain is \emph{recurrent}, if it holds $\mathbb{E}\left\{
\eta_{\mathcal{A}}\right\}  =+\infty$, $\forall x \in\mathcal{A}$ for any set
$\mathcal{A} \in\mathbb{R}^{M}_{+}$. Additionally, if the Markov chain admits
an invariant probability measure $\pi$, then it is \emph{positive recurrent}.
\end{definition}

If the Markov chain is positive recurrent, it is also \emph{weakly stable}
\cite{Leonardi:2001} so that it holds
\[
\lim_{n\rightarrow+\infty} Pr(\|\mathbf{q}(n)\|>B) < \epsilon
\]
for any $\epsilon>0$ and some constant $B>0$.

\begin{definition}
A Markov chain is \emph{transient}, if there is a countable cover of
$\mathbb{R}_{+}^{M}$ with uniformly transient sets, i.e. there is a constant
$C$ with $\mathbb{E}\left\{  \eta_{\mathcal{A}}\right\}  \leq C$, $\forall x
\in\mathcal{A}$.
\end{definition}

In this paper we also apply another stability definition as it is used in
\cite{Eryilmaz:2005}:

\begin{definition}
\label{definition_f_stable} A Markov chain is called \emph{f-stable}, if there
is an unbounded function $f:\mathbb{R}^{M}_{+} \rightarrow\mathbb{R}_{+}$ so
that for any $0<B<+\infty$ the set $\mathcal{B}:=\left\{  \mathbf{x}:f\left(
\mathbf{x}\right)  \leq B\right\}  $\ is compact, and furthermore it holds
\begin{equation}
\label{condition_f_stable}\limsup_{n\rightarrow+\infty}\mathbb{E}\left\{
f\left(  \mathbf{q}\left(  n\right)  \right)  \right\}  <+\infty.
\end{equation}

\end{definition}

The function $f$ is unbounded in all positive directions so that $f\left(
\mathbf{q}\left(  n\right)  \right)  $ goes to infinity when $\|\mathbf{q}\| $
goes to infinity. Choosing directly $f\left(  \mathbf{q}\right)
=\|\mathbf{q}\|$, Definition \ref{definition_f_stable} is equivalent to the
definition of \emph{strongly stable}\cite{Leonardi:2001}. Moreover, it is easy
to show that for any $f\left(  \mathbf{q}\right)  $ which grows faster than
$\Vert\mathbf{q}\Vert$, inequality (\ref{condition_f_stable}) implies that the
Markov chain is strongly stable.

Denoting the mean of the arrival bits $a_{i}(n)$ per time slot as $\rho_{i}$
collected in the vector $\boldsymbol{\rho}\in\mathbb{R}_{+}^{M}$, we call a
vector of arrival rates $\boldsymbol{\rho}$ \emph{stabilizable }under\emph{
}$\mathcal{P}$ when the corresponding queueing system driven by some specific
scheduling policy $\mathcal{P}$ is positive recurrent.

It is well-known that any vector of arrival rates inside the ergodic
achievable rate region $\mathcal{C}(\hat{P})$ is stabilizable (e.g. under MWM
policy) and any vector of arrival rate outside $\mathcal{C}(\hat{P})$ is not
stabilizable \cite{Eryilmaz:2005,Wireless_Exponential}. Thus a scheduling
policy is now called \emph{throughput-optimal} if it keeps the Markov chain
positive recurrent for any vector of arrival rates $\boldsymbol{\rho}%
\in\text{int}(\mathcal{C}(\hat{P}))$, where $\text{int}(\mathcal{C}(\hat{P}))$
denotes the interior of the ergodic achievable rate region $\mathcal{C}%
(\hat{P})$.


\section{Stability Conditions}

\label{section:stability}

We consider scheduling policies which solve the weighted sum rate maximization
problem
\begin{equation}
\mathbf{r}^{\mathcal{P}}(\mathbf{h}(n),\mathbf{q}(n))=\underset{\tilde
{\mathbf{r}}\in\mathcal{C}\left(  \mathbf{h}(n),\hat{P}\right)  }{\arg\max
}\boldsymbol{\mu}^{\mathcal{P}}\left(  \mathbf{q}(n)\right)  \tilde
{\mathbf{r}}, \label{eqn:weighted_sum_rate_3}%
\end{equation}
where $\boldsymbol{\mu}^{\mathcal{P}}\left(  \mathbf{q}\right)  $
denotes the weight vector for some queue state $\mathbf{q}$
determined by a scheduling policy $\mathcal{P}$. Note that the
weight factors $\boldsymbol{\mu }^{\mathcal{P}}\left(
\mathbf{q}\right)  $ depends solely on the queue state. In Section
\ref{section:necessity} we argue for the necessity of this
assumption. It is worth noting that the solution of the
optimization problem in (\ref{eqn:weighted_sum_rate_3}) is a
boundary point of the convex hull of
$\mathcal{C}(\mathbf{h}(n),\hat{P})$. However, such a rate
allocation is possibly not uniquely defined by the weight vector
$\boldsymbol{\mu}^{\mathcal{P}}( \mathbf{q})$ (which is often the
case). Nevertheless we can enforce uniqueness by invoking e.g.
additional constraints on the allocated rate vector which, by the
way, do not affect the line of proof in Theorem
\ref{theorem:convergence}.

One of the well-known throughput-optimal scheduling policies is the MWM
policy, which uses the weight vector
\begin{equation}
\boldsymbol{\mu}^{MWM}(\mathbf{q})=\mathbf{q}. \label{weight_MWM}%
\end{equation}
The throughput-optimality of the MWM policy in general multiple-access and
broadcast channels is proven in \cite{NeeModRoh:2003, YehCoh:2003}. Despite
its simple form the MWM policy has satisfactory delay and fairness properties
and is applied in several systems including MIMO
\cite{Viswanathan_Kumaran_allerton01} and OFDM.

A further class of throughput-optimal scheduling policies use the exponential
rule \cite{Shakkottai:2001}. Here, the weight vector is given by
\begin{equation}
\mu_{i}^{EXP}(\mathbf{q})=\gamma_{i} exp\left(  \frac{\alpha_{i}q_{i}}%
{\beta+\left(  \frac{1}{M}\sum_{j\in\mathcal{M}}\alpha_{j}q_{j}\right)
^{\eta}}\right)  , \label{weight_exp_rule}%
\end{equation}
where $\gamma_{1},...,\gamma_{M}$, $\alpha_{1},...,\alpha_{M}$ are arbitrary
sets of positive constants and the positive constants $\beta$ and $\eta
\in(0,1)$ are fixed.

A more generalized class of throughput-optimal scheduling policies is
presented by Eryilmaz et al. in \cite{Eryilmaz:2005}. The weight factor
$\mu_{i}(q_{i})$ is given by a function of $q_{i}$ satisfying the following conditions:

\begin{enumerate}
\item $\mu_{i}(q_{i})$ is a nondecreasing, continuous function with
$\lim_{q_{i}\rightarrow+\infty}\mu_{i}(q_{i})=+\infty$.

\item Given any $C_{1}>0$, $C_{2}>0$ and $0<\epsilon<1$ there exists a
$B<+\infty$, such that for all $q_{i}>B$ and $\forall i\in\mathcal{M}$,
\begin{align}
(1-\epsilon)\mu_{i}(q_{i})  &  \leq\mu_{i}(q_{i}-C_{1})\nonumber\\
&  \leq\mu_{i}(q_{i}+C_{2})\leq(1+\epsilon)\mu_{i}(q_{i}).
\label{condition_Eryilmaz}%
\end{align}

\end{enumerate}

Condition (\ref{condition_Eryilmaz}) implies that the relative difference
$\frac{\left\vert \mu_{i}(q_{i}\pm C)-\mu_{i}(q_{i})\right\vert }{\mu
_{i}(q_{i})}$ tends toward zero for constant $C$ if $q_{i}$ is large. Hence
the scheduling policies using weight functions such as $\mu_{i}^{\mathcal{P}%
}=e^{q_{i}}$ do not belong to this class. Actually, it can be proven that
these scheduling policies are not throughput-optimal.

The conditions given in \cite{Eryilmaz:2005} cover quite a large class of
throughput-optimal scheduling policies. However, the weight factor $\mu
_{i}(q_{i})$ is exclusively calculated by $q_{i}$ and independent of the queue
length of other users, which is a rather strict constraint. In general the
weight factor is determined by the queue state of all users, i.e.
\[
\mu_{i}:\mathbb{R}_{+}^{M}\rightarrow\mathbb{R}_{+}, \mathbf{q} \mapsto\mu
_{i}(\mathbf{q}).
\]
Some examples are the aforementioned policies using exponential rule, the QPS
and the ISPS. Unfortunately the results in \cite{Eryilmaz:2005} can not be
applied in these cases.

Here, we give generalized sufficient conditions for throughput-optimality. The
conditions are presented by characterizing the corresponding weight vector
$\boldsymbol{\mu}^{\mathcal{P}}$ of the scheduling policies. In the following
we consider the normalized weight vector
\begin{equation}
\boldsymbol{\bar{\mu}}^{\mathcal{P}}(\mathbf{q}) := \frac{\boldsymbol{\mu
}^{\mathcal{P}}(\mathbf{q})}{\left\Vert \boldsymbol{\mu}^{\mathcal{P}}(
\mathbf{q})\right\Vert _{1}} \label{eqn:normalize}%
\end{equation}
and hence $\left\Vert \boldsymbol{\bar{\mu}}^{\mathcal{P}}( \mathbf{q}%
)\right\Vert _{1}= 1$. Since the magnitude of the weight vector does not
affect the solution of the maximization problem
(\ref{eqn:weighted_sum_rate_ins}), namely the scheduling decision, we only
need to consider the direction of the vector. Thus the normalization of weight
factors $\boldsymbol{\mu}^{\mathcal{P}}(\mathbf{q})$ can be done without lose
of generality.

\begin{theorem}
Any vector of arrival rates $\boldsymbol{\rho}\in\text{int} (\mathcal{C}(
\hat{P})) $ is stabilizable under the scheduling policy $\mathcal{P}$, if its
corresponding normalized weight vector $\boldsymbol{\bar{\mu}}(\mathbf{q})$
given in Eqn.(\ref{eqn:normalize}) fulfills the following conditions:

\begin{enumerate}
\item Given any $0<\epsilon_{1}<1$ and $C_{1}>0$, there is some $B_{1}>0$ so
that for any $\Delta\mathbf{q\in}\mathbb{R}^{M}$ with $\left\Vert
\Delta\mathbf{q}\right\Vert \mathbf{<}C_{1}$, we have $\left\vert \bar{\mu
}_{i}\left(  \mathbf{q}+\Delta\mathbf{q}\right)  -\bar{\mu}_{i}\left(
\mathbf{q}\right)  \right\vert \leq\epsilon_{1}$ for any $\mathbf{q}%
\in\mathbb{R}_{+}^{M}$ with $\left\Vert \mathbf{q}\right\Vert >B_{1}$,
$\forall i\in\mathcal{M}$.

\item Given any $0<\epsilon_{2}<1$ and $C_{2}>0$, there is some $B_{2}>0$ so
that for any $\mathbf{q}\in\mathbb{R}_{+}^{M}$ with $\left\Vert \mathbf{q}%
\right\Vert >B_{2}$ and $q_{i}<C_{2}$, we have $\bar{\mu}_{i}(\mathbf{q}%
)\leq\epsilon_{2}$, for any $i\in\mathcal{M}$.
\end{enumerate}

Moreover, for any arrival process with $\boldsymbol{\rho}\in\text{int}
(\mathcal{C}( \hat{P})) $, the queueing system is f-stable under the given
policy $\mathcal{P}$, where $f$ is an unbounded function as defined in
Definition \ref{definition_f_stable}. The exact formulation of $f$ depends on
the weight function $\boldsymbol{\bar{\mu}}(\mathbf{q})$.

\label{theorem:convergence}
\end{theorem}

In order to simplify the notation, provided the limit exists, we can also
write the two conditions as
\begin{align*}
&  \lim_{\left\Vert \mathbf{q}\right\Vert \rightarrow+\infty}\left\vert
\bar{\mu}_{i}\left(  \mathbf{q}+\Delta\mathbf{q}\right)  -\bar{\mu}_{i}\left(
\mathbf{q}\right)  \right\vert =0, & \left\Vert \Delta\mathbf{q}\right\Vert
\mathbf{<}C_{1}  & \\
&  \lim_{\left\Vert \mathbf{q}\right\Vert \rightarrow+\infty}\left\vert
\bar{\mu}_{i}\left(  \mathbf{q}\right)  \right\vert =0, & q_{i}<C_{2}  & ,
\end{align*}
where $\| \mathbf{q}\| \rightarrow+\infty$ is any path in $\mathbb{R}^{M}_{+}$
with unbounded norm. The conditions are interpreted in the next section and
their necessity is proven in Theorem \ref{theorem:necessity}. Before we give a
proof of Theorem \ref{theorem:convergence}, we compare the conditions in
Theorem \ref{theorem:convergence} and the conditions given by Eryilmaz et al.
To this end, let $\mu_{i}(q_{i})$ be the function dependent only on $q_{i}$
and satisfies Eryilmaz's conditions. We consider two different cases:
$q_{i}<+\infty$ and $q_{i}\rightarrow+\infty$. If $q_{i}$ is bounded and
$\left\Vert \mathbf{q}\right\Vert $ goes to infinity, then we have some $j\neq
i$ with $q_{j}\rightarrow+\infty$. According to Eryilmaz's conditions, it
follows that $\mu_{i}(q_{i})<+\infty$ and $\mu_{j}(q_{j})\rightarrow+\infty$.
We normalize the weight vector so that we have
\begin{align*}
&  \lim_{\left\Vert \mathbf{q}\right\Vert \rightarrow+\infty}\bar{\mu}%
_{i}(\mathbf{q})\\
=  &  \lim_{q_{j}\rightarrow+\infty}\frac{\mu_{i}(q_{i})}{\mu_{j}(q_{j}%
)+\sum_{k\neq j}\mu_{k}(q_{k})}=0
\end{align*}
and Condition 2) in Theorem \ref{theorem:convergence} is fulfilled. It holds
also that $\lim_{\left\Vert \mathbf{q}\right\Vert \rightarrow+\infty}\bar{\mu
}_{i}(\mathbf{q}+\Delta\mathbf{q})=0$ as long as $\left\Vert \Delta
\mathbf{q}\right\Vert $ is bounded. Thus $\lim_{\left\Vert \mathbf{q}%
\right\Vert \rightarrow+\infty}\left\vert \bar{\mu}_{i}(\mathbf{q}%
+\Delta\mathbf{q})-\bar{\mu}_{i}(\mathbf{q})\right\vert =0$ and Condition 1)
is satisfied.

If $q_{i}\rightarrow+\infty$ we only need to check the first condition in
Theorem \ref{theorem:convergence}. Suppose $\left\Vert \Delta\mathbf{q}%
\right\Vert <C$ for some constant $C>0$, we have $\Delta q_{i}<C$, $\forall
i\in\mathcal{M}$. After the normalization we have
\begin{equation}
\bar{\mu}_{i}(\mathbf{q}+\Delta\mathbf{q})=\frac{1}{1+\sum_{j\neq i}\frac
{\mu_{j}(q_{j}+\Delta q_{j})}{\mu_{i}(q_{i}+\Delta q_{i})}}.
\label{eqn:normalize_dq}%
\end{equation}
As $\left\Vert \mathbf{q}\right\Vert \rightarrow+\infty$, if there are other
users $j\neq i$ with $q_{j}\rightarrow+\infty$, then according to
(\ref{condition_Eryilmaz}) we have
\[
\lim_{\left\Vert \mathbf{q}\right\Vert \rightarrow+\infty}\frac{\mu_{j}%
(q_{j}+\Delta q_{j})}{\mu_{i}(q_{i}+\Delta q_{i})}=\lim_{\left\Vert
\mathbf{q}\right\Vert \rightarrow+\infty}\frac{\mu_{j}(q_{j})}{\mu_{i}(q_{i}%
)}.
\]
Otherwise if $q_{j}$ is bounded, it holds
\[
\lim_{\left\Vert \mathbf{q}\right\Vert \rightarrow+\infty}\frac{\mu_{j}%
(q_{j}+\Delta q_{j})}{\mu_{i}(q_{i}+\Delta q_{i})}=0.
\]
Considering the both situations and substitute them in (\ref{eqn:normalize_dq}%
), we have $\lim_{\left\Vert \mathbf{q}\right\Vert \rightarrow+\infty
}\left\vert \bar{\mu}_{i}(\mathbf{q}+\Delta\mathbf{q})-\bar{\mu}%
_{i}(\mathbf{q})\right\vert =0$ and the condition is satisfied. Hence we
conclude that the class of policies in \cite{Eryilmaz:2005} is indeed included
in the theorem.

\begin{proof}
The proof is given in Appendix A.
\end{proof}

So far we considered scheduling policies based on the current queue state. In
some situations the queue state information might be imprecise or delayed.
These cases occur even more frequently in an uplink system, where the queue
state information has to be quantized and transmitted from the mobile terminal
to the base station through a signaling channel. In this paper we mainly
consider the downlink system, however we emphasize that the physical layer
described in Section \ref{section:system} can be generalized as the achievable
rate region is independent of the transmission schemes. Thus our results can
also be applied to multiple-access channels in uplink systems if we replace
the term downlink achievable rate region with uplink achievable rate region in
our system model.

Theorem \ref{theorem:convergence} has an interesting interpretation: suppose
the scheduling policy determines the rate allocation based on the quantized
queue state information $\mathbf{\bar{q}}(n)$ with some quantization error
$\boldsymbol{\varepsilon}(\mathbf{q}(n))$, then we have $\mathbf{\bar{q}%
}(n)=\mathbf{q}(n)+\boldsymbol{\varepsilon}(\mathbf{q}(n))$. If the weight
vector $\bar{\boldsymbol{\mu}}(\mathbf{\bar{q}}(n))$ determined by the
scheduling policy satisfies the conditions given in Theorem
\ref{theorem:convergence} when $\Vert\mathbf{\bar{q}}(n)\Vert$ is sufficiently
large, it is easy to show that
\[
\left\vert \bar{\mu}_{i}\left(  \mathbf{q}+\Delta\mathbf{q+}%
\boldsymbol{\varepsilon}\left(  \mathbf{q}+\Delta\mathbf{q}\right)  \right)
-\bar{\mu}_{i}\left(  \mathbf{q+}\boldsymbol{\varepsilon}\left(
\mathbf{q}\right)  \right)  \right\vert \leq\epsilon_{1}%
\]
for any $0<\epsilon_{1}<1$ and bounded $\Delta\mathbf{q}$, and if $q_{i}$ is
bounded, it holds%
\[
\bar{\mu}_{i}(\mathbf{q+}\boldsymbol{\varepsilon}\left(  \mathbf{q}\right)
)\leq\epsilon_{2}%
\]
for any $0<\epsilon_{2}<1$ as $\Vert\mathbf{q}(n)\Vert$ is sufficiently large.
Thus the scheduling policy is also throughput-optimal.

If the obtained queue state information has $\Delta n$ time slots delay, we
have $\mathbf{\bar{q}}(n)=\mathbf{q}(n-\Delta n)$. Since the transmission and
arrival rate are bounded, we have $\mathbf{\bar{q}}(n)=\mathbf{q}%
(n)+\boldsymbol{\varepsilon}_{d}$ where the error $\boldsymbol{\varepsilon
}_{d}$ caused by delay is also bounded. Similarly it can be shown that the
stability conditions can also be applied in this case.

\section{Necessity of Stability Conditions}

\label{section:necessity}

In the previous section we presented sufficient conditions for
throughput-optimal scheduling policies. To do so, the scheduling problem is
formulated as a weighted sum rate maximization problem, and, hence, the rate
allocation of these scheduling policies always lies on the convex hull of the
instantaneous achievable rate region $\mathcal{C}(\mathbf{h}(n),\hat{P})$.
Additionally, the weight factors $\boldsymbol{\mu}^{\mathcal{P}}$ are
independent of the instantaneous channel state. Actually, we can observe that
all existing throughput-optimal policies have these general characters
(although it is not explicitly noticed in previous works). In fact it is an
inherent necessity of any throughput-optimal scheduling policy which is
expressed as the following theorem.

\begin{theorem}
\label{theorem:weighted_R} If the queue lengths are sufficiently large,

\begin{enumerate}
\item a throughput-optimal policy always allocates the rate vector on the
convex hull of the instantaneous rate region thus the rate allocation can be
formulated as a weighted sum rate maximization problem.

\item Furthermore, the weight vector $\boldsymbol{\mu}^{\mathcal{P}}$ in the
maximization problem is independent of the current fading state $\mathbf{h}%
(n)$.
\end{enumerate}
\end{theorem}

The theorem is proven in \cite{EURASIP_ISP}. Next we consider the
necessity of the stability conditions given in Theorem
\ref{theorem:convergence}. It is immediately clear that the
conditions are not universally necessary. In some specific
scenario the achievable rate region has no unique supporting
hyperplane for some point on the boundary. Hence, the weight
vector on these points is not unique. A typical example is a rate
region with only two available rate points on the boundary. In
this case a throughput-optimal scheduling policy can be
characterized by a weight function where the image consists of two
points in $\mathbb{R}^M_+$ only. Obviously this weight function
does not satisfy Condition 1) in Theorem
\ref{theorem:convergence}. To fix this problem, we argue that in
the wireless case the achievable rate regions are varying over
time and further the policies must be defined for any possible
configuration of $\mathcal{C}(\mathbf{h}(n),\hat{P})$. Hence, to
prove the necessity we will only considered those achievable rate
regions where the weight vector is unique for every boundary
point.

Before we prove the necessity under above assumptions, we want to
give some intuition about why these conditions must hold in
general. Recall that a throughput-optimal policy should keep the
queues stable for any mean arrival rate $\boldsymbol{\rho}$ inside
the ergodic rate region $\mathcal{C}(\hat{P})$; so if the arrival
rate vector $\boldsymbol{\rho}$ lies close to some boundary point
$\mathbf{r}^{\ast}$ of $\mathcal{C}(\hat{P})$ corresponding to a
weight vector $\boldsymbol{\bar{\mu}}^{\ast}$, heuristically, the
weight vector determined by a throughput-optimal scheduling
$\boldsymbol{\bar{\mu}}^{\mathcal{P}}$ should be in the close
neighborhood of $\boldsymbol{\bar{\mu}}^{\ast}$ for almost all
time slots. Condition 1) in Theorem \ref{theorem:convergence}
ensures now that the weight vector varies smoothly between two
time slots if the queue lengths become large. Thus for the above
situation, it guarantees that the weight vector $\boldsymbol{\bar
{\mu}}^{\mathcal{P}}$ does not leave the neighborhood of
$\boldsymbol{\bar {\mu}}^{\ast}$ in almost all time slots.
Condition 2) in Theorem \ref{theorem:convergence} guarantees that
no rate is wasted on "nonurgent" users. If the queues of some
users are bounded while the other queues expand, the scheduler
should reduce the weights on these users and save these rate
resources for other users.

\begin{theorem}
\label{theorem:necessity} A scheduling policy $\mathcal{P}$ is not
throughput-optimal, namely there exists some arrival process with
$\boldsymbol{\rho}\in\text{int} (\mathcal{C}(\hat{P}))$ which is not
stabilizable under the policy $\mathcal{P}$, if the policy has one of the
following characteristics:

\begin{enumerate}
\item The change of the weight vector between two time slots is
not negligible, i.e., there is some constant $0<\gamma\leq1$ and
$\epsilon>0$ so that it holds
\begin{align}
&  \lim_{N\rightarrow+\infty}\frac{1}{N}\sum_{n=1}^{N}\mathbb{I}\left\{
\left\Vert \boldsymbol{\bar{\mu}}\left(  \mathbf{q}(n+1)\right)
-\boldsymbol{\bar{\mu}}\left(  \mathbf{q}(n)\right)  \right\Vert \geq
\epsilon\right\}  \geq\gamma\label{condition_necessity_C1_compl}%
\end{align}
for any $\mathbf{q}(n)$, $n \in \mathbb{N}$, with probability $1$.

\item There is some user $i\in\mathcal{M}$, whose weight factor is
not negligible, i.e., there is some constant $0<\gamma\leq1$ and
$\epsilon>0$ so that it holds
\begin{equation}
\lim_{N\rightarrow+\infty}\frac{1}{N}\sum_{n=1}^{N}\mathbb{I}\left\{  \bar
{\mu}_{i}\left(  \mathbf{q}(n)\right)  \geq\epsilon\right\}  \geq
\gamma\label{condition_necessity_C2_compl}%
\end{equation}
for any $\mathbf{q}(n)$, $n \in \mathbb{N}$, with probability $1$.

\end{enumerate}
\end{theorem}

\begin{proof}
The proof is given in Appendix B.
\end{proof}

Applying the results in Theorem \ref{theorem:necessity}, it can be
proven that the policies using certain exponential functions as
weight factors are not throughput-optimal. The details are given
in Section \ref{section:applications}.

\section{Applications}

\label{section:applications}

In this section we prove the throughput-optimality of some well-known
scheduling policies. Note that the throughput-optimality of these policies are
already proven in the previous works, here we use our results to perform the
proof in a different way and show the applicability of our results. In this
section we also use the necessary conditions in Theorem
\ref{theorem:necessity} and verify some policies which are not throughput-optimal.

Normalizing the weight factors given in (\ref{weight_MWM}), it is
easy to show that the MWM policy satisfies the conditions in
Section \ref{section:stability}, and, hence, it is
throughput-optimal. In the following we use the results to check
the throughput-optimality of several other scheduling policies.

\subsection{Exponential Rule}

Normalizing the weight factors given in (\ref{weight_exp_rule}), we have%
\begin{align}
\bar{\mu}_{i}(\mathbf{q})  &  =\frac{\gamma_{i}}{\sum_{j\in\mathcal{M}}%
\gamma_{j}exp\left(  \frac{\alpha_{j}q_{j}-\alpha_{i}q_{i}}{\beta+\left(
\frac{1}{M}\sum_{k\in\mathcal{M}}\alpha_{k}q_{k}\right)  ^{\eta}}\right)
}\nonumber\\
&  =\frac{\gamma_{i}}{\sum_{j\in\mathcal{M}}\gamma_{j}exp\left(  \frac
{\alpha_{j}q_{j}-\alpha_{i}q_{i}}{\beta+\widetilde{\alpha}\left(
\mathbf{q}\right)  \left\Vert \mathbf{q}\right\Vert _{2}^{\eta}}\right)  }.
\label{eqn:exponential_normal}%
\end{align}
Recall that $\gamma_{1},...,\gamma_{M}$, $\alpha_{1},...,\alpha_{M}$, $\beta$,
$\eta$ are predefined constants in the scheduler and we define $\widetilde
{\alpha}\left(  \mathbf{q}\right)  :=\left(  \frac{1}{M}\cdot\frac
{\boldsymbol{\alpha}^{T}\mathbf{q}}{\left\Vert \boldsymbol{\alpha}\right\Vert
_{2}\left\Vert \mathbf{q}\right\Vert _{2}}\right)  ^{\eta}$.

In order to check the first condition in Theorem
\ref{theorem:convergence}, we need to show that
\begin{equation}
\lim_{\left\Vert \mathbf{q}\right\Vert \rightarrow+\infty}\left\vert \bar{\mu
}_{i}\left(  \mathbf{q}+\Delta\mathbf{q}\right)  -\bar{\mu}_{i}\left(
\mathbf{q}\right)  \right\vert =0. \label{eqn:exponential_c1}%
\end{equation}
If $\left\Vert \Delta\mathbf{q}\right\Vert $ is bounded, it holds
$\lim_{\left\Vert \mathbf{q}\right\Vert \rightarrow+\infty}\widetilde{\alpha
}\left(  \mathbf{q}+\Delta\mathbf{q}\right)  =\lim_{\left\Vert \mathbf{q}%
\right\Vert \rightarrow+\infty}\widetilde{\alpha}\left(  \mathbf{q}\right)  $
and%
\begin{align*}
&  \lim_{\left\Vert \mathbf{q}\right\Vert \rightarrow+\infty}\frac{\alpha
_{j}\left(  q_{j}-\Delta q_{j}\right)  -\alpha_{i}\left(  q_{i}-\Delta
q_{i}\right)  }{\beta+\widetilde{\alpha}\left(  \mathbf{q}+\Delta
\mathbf{q}\right)  \left\Vert \mathbf{q}+\Delta\mathbf{q}\right\Vert
_{2}^{\eta}}\\
=  &  \lim_{\left\Vert \mathbf{q}\right\Vert \rightarrow+\infty}\frac
{\alpha_{j}q_{j}-\alpha_{i}q_{i}}{\beta+\widetilde{\alpha}\left(
\mathbf{q}\right)  \left\Vert \mathbf{q}\right\Vert _{2}^{\eta}}%
\end{align*}
for all $i,j\in\mathcal{M}$ so that the Eqn. (\ref{eqn:exponential_c1}) follows.

Considering the second condition in Theorem \ref{theorem:convergence}, if
$q_{i}$ is bounded as $\left\Vert \mathbf{q}\right\Vert $ increases, there is
another user $j$\ who has the longest queue so that $q_{j}\geq\frac{\left\Vert
\mathbf{q}\right\Vert }{M}$, then we have%
\begin{equation}
\lim_{\left\Vert \mathbf{q}\right\Vert \rightarrow+\infty}\frac{\alpha
_{j}q_{j}-\alpha_{i}q_{i}}{\beta+\widetilde{\alpha}\left(  \mathbf{q}\right)
\left\Vert \mathbf{q}\right\Vert _{2}^{\eta}}=\lim_{\left\Vert \mathbf{q}%
\right\Vert \rightarrow+\infty}\frac{\alpha_{j}q_{j}}{\widetilde{\alpha
}\left(  \mathbf{q}\right)  \left\Vert \mathbf{q}\right\Vert _{2}^{\eta}%
}=+\infty\label{eqn:exponential_l2}%
\end{equation}
since $\eta<1$.

Substituting (\ref{eqn:exponential_l2}) in
Eqn.(\ref{eqn:exponential_normal}),
it follows%
\[
\lim_{\left\Vert \mathbf{q}\right\Vert \rightarrow+\infty}\bar{\mu}_{i}\left(
\mathbf{q}\right)  =0
\]
which fulfills the Condition 2) in Theorem \ref{theorem:convergence} and the
throughput-optimality is proven.

\subsection{QPS}

QPS from \cite{Seong:2006} is a scheduling policy which has good
delay and fairness performance in the downlink. Applying QPS in a
broadcast system with random arrivals, each user's queueing delay
becomes equal as $n\rightarrow+\infty$. Additionally, if the queue
state is initialized by $\mathbf{q}\left(  0\right)
>0$ and there is no new packet arrivals after $n=0$, which can be
considered as a draining problem, the QPS minimizes the expected
draining time until all the buffers are cleared.

\begin{figure}[h]
\begin{center}
\psfrag{boundary of Cxxxxxxxx}{\tiny boundary of $\mathcal{C}(
\hat{P}) $} \psfrag{q}{$\mathbf{q}$} \psfrag{ua}{$\boldsymbol{\mu}^{MWM}$}
\psfrag{ub}{$\boldsymbol{\mu}^{QPS}$}
\includegraphics[width=1\linewidth]{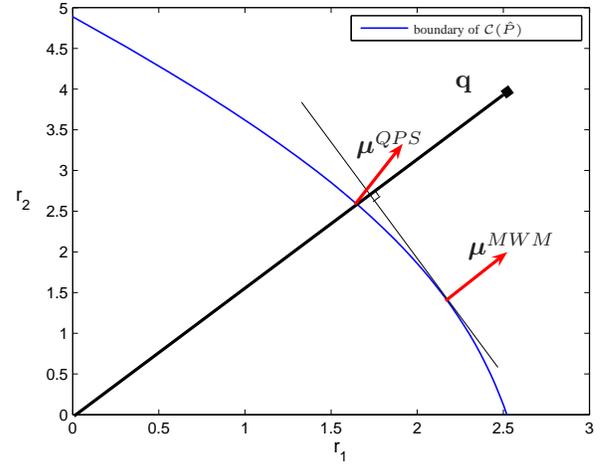}
\end{center}
\par
.\caption{The weight vector of QPS and MWM policy. For MWM $\boldsymbol{\mu
}^{MWM} = \mathbf{q}$ and for QPS the weight vector $\boldsymbol{\mu}^{QPS}$
is the norm at the boundary point of $\mathcal{C}\left(  \hat{P}\right)  $
where \ $\mathbb{E}\left\{  \left.  \mathbf{r}^{\mathcal{P}}\left(
\mathbf{h},\mathbf{q}\right)  \right\vert \mathbf{q}\right\}  $ is
proportional to $\mathbf{q}$}%
\label{fig_rate_region_QPS}%
\end{figure}

The rate vector is allocated so that
\[
\mathbb{E}\left\{  \mathbf{r}^{\mathcal{P}}\left(  \mathbf{h}\left(  n\right)
,\mathbf{q}\left(  n\right)  \right)  \right\}  =\mathbf{q}\left(  n\right)
\underset{x\mathbf{q}^{n}\in\mathcal{C}\left(  \hat{P}\right)  }{\max}x,
\]
where $x$ is a scalar. According to the policy, the weight vector is chosen as
the norm at the boundary point of $\mathcal{C}( \hat{P}) $ where
\ $\mathbb{E}\left\{  \left.  \mathbf{r}^{\mathcal{P}}\left(  \mathbf{h}%
,\mathbf{q}\right)  \right\vert \mathbf{q}\right\}  $ is proportional to
$\mathbf{q}$. Fig.\ref{fig_rate_region_QPS} shows the expected rate vector
allocated by QPS compared to MWM policy and the ergodic achievable rate region
in a 2-user scenario.

The weight vector in QPS is found by \emph{Geometric Programming}.
Unfortunately, there exists no explicit function of the weight
factor $\boldsymbol{\mu}$ and the rate allocation
$\mathbf{r}^{\mathcal{P}}$ in QPS. Therefore the verification of
its throughput-optimality is not easy using standard methods. The
presented general approach in this paper is particularly suitable
in this case.

Since the weight vector is determined by the normalized queue state
$\frac{\mathbf{q}}{\left\Vert \mathbf{q}\right\Vert _{2}}$, for bounded
$\left\Vert \Delta\mathbf{q}\right\Vert $ we have
\begin{align*}
&  \lim_{\left\Vert \mathbf{q}\right\Vert \rightarrow+\infty}\bar{\mu}%
_{i}(\mathbf{q}+\Delta\mathbf{q})\\
=  &  \lim_{\left\Vert \mathbf{q}\right\Vert \rightarrow+\infty}\bar{\mu}%
_{i}(\frac{\mathbf{q}+\Delta\mathbf{q}}{\left\Vert \mathbf{q}\right\Vert _{2}%
})\\
=  &  \lim_{\left\Vert \mathbf{q}\right\Vert \rightarrow+\infty}\bar{\mu}%
_{i}(\mathbf{q)}.
\end{align*}

If $q_{i}<+\infty$ we have
\[
\lim_{\left\Vert \mathbf{q}\right\Vert \rightarrow+\infty}\frac{q_{i}%
}{\left\Vert \mathbf{q}\right\Vert _{2}}=0
\]
and hence
\[
\lim_{\left\Vert \mathbf{q}\right\Vert \rightarrow+\infty}\bar{\mu}%
_{i}(\mathbf{q)}=0.
\]

Both conditions in Theorem \ref{theorem:convergence} are satisfied and the QPS
policy is throughput-optimal.

\subsection{ISPS}

Another scheduling policy which also provides superior delay performance is
the ISPS in \cite{EURASIP_ISP}. If there is no new packet arrivals in the
system and the queue state is initialized by $\mathbf{q}^{0}>0$, the policy
minimizes the average queueing delay
\begin{equation}
\bar{D}(N):=\frac{1}{M}\sum_{i=1}^{M}D_{i}(N)=\frac{1}{MN}\sum_{i=1}^{M}%
\sum_{n=1}^{N}\frac{q_{i}\left(  n\right)  }{\mathbb{E}\left\{  a_{i}\left(
n\right)  \right\}  }, \label{eqn:D}%
\end{equation}
where $N$ is the length of the observed time slots.

The weight factor $\bar{\mu}_{i}(\mathbf{q})$ is determined by the estimated
service time $\eta_{i}$ such that
\begin{equation}
\bar{\mu}_{i}(\mathbf{q})=\frac{\frac{\eta_{i}(\mathbf{q})}{\overline{a}_{i}}%
}{\sum_{j\in\mathcal{M}}\frac{\eta_{j}(\mathbf{q})}{\overline{a}_{j}}},
\end{equation}
where $\overline{a}_{i},\overline{a}_{j}\in\mathbb{R}_{+}$ are some predefined
constants. Parameter $\eta_{i}(\mathbf{q})$ is the expected service time of
user $i$ if the ISPS is applied in the system and no new packet arrival
occurs. Then in the time slot $n>\eta_{i}(\mathbf{q})$ the buffer of user $i$
is completely emptied and the corresponding transmitter is in idle state. The
estimated service time $\boldsymbol{\eta}$ \ is obtained with an iterative
algorithm. Hence there is also no explicit function of $\mu_{i}(\mathbf{q})$
in ISPS.

However, we can see that if $\left\Vert \Delta\mathbf{ q}\right\Vert <+\infty
$, it follows $\Delta\eta_{i}:=\left\vert \eta_{i}(\mathbf{q}+\Delta
\mathbf{q})-\eta_{i}(\mathbf{q})\right\vert <+\infty$. If $q_{i}%
\rightarrow+\infty$, then $\eta_{i}(\mathbf{q})\rightarrow+\infty$. It holds
\begin{align*}
&  \lim_{\left\Vert \mathbf{q}\right\Vert \rightarrow+\infty}\left\vert
\bar{\mu}_{i}\left(  \mathbf{q}+\Delta\mathbf{q}\right)  -\bar{\mu}_{i}\left(
\mathbf{q}\right)  \right\vert \\
=  &  \lim_{\left\Vert \mathbf{q}\right\Vert \rightarrow+\infty}\left\vert
\frac{\frac{\Delta\eta_{i}}{\overline{a}_{i}}}{\frac{\eta_{i}(\mathbf{q}%
)}{\overline{a}_{i}}+\sum_{j\neq i}\frac{\eta_{j}(\mathbf{q})}{\overline
{a}_{j}}}\right\vert =0.
\end{align*}

If $q_{i}$ is bounded we have $\eta_{i}(\mathbf{q})<+\infty$ and
\[
\lim_{\left\Vert \mathbf{q}\right\Vert \rightarrow+\infty}\sum_{j\neq i}%
\eta_{j}(\mathbf{q})=+\infty
\]
and
\[
\lim_{\left\Vert \mathbf{q}\right\Vert \rightarrow+\infty}\bar{\mu}%
_{i}(\mathbf{q})=\lim_{\left\Vert \mathbf{q}\right\Vert \rightarrow+\infty
}\frac{\frac{\eta_{i}(\mathbf{q})}{\overline{a}_{i}}}{\frac{\eta
_{i}(\mathbf{q})}{\overline{a}_{i}}+\sum_{j\neq i}\frac{\eta_{j}(\mathbf{q}%
)}{\overline{a}_{j}}}=0.
\]

Thus the throughput-optimality is proven.

\subsection{Exponential Functions as Weight factors}

Applying Theorem \ref{theorem:necessity}, it can easily be shown that the
policy using exponential weight function such as $\mu_{i}^{\mathcal{P}%
}(\mathbf{q})=e^{q_{i}}$ is not throughput-optimal. We consider a point
$\mathbf{r}^{\ast}$ on the boundary of ergodic achievable rate region
$\mathcal{C}(\hat{P})$ and its corresponding normal vector $\boldsymbol{\mu
}^{\ast}$ with $\mu_{i}^{\ast}>0$, $\forall i\in\mathcal{M}$. Suppose the
expected arrival rate vector $\boldsymbol{\rho}$ lies close to the boundary
point $\mathbf{r}^{\ast}$, in order to keep the system stable, the weight
factors $\bar{\boldsymbol{\mu}}$ should be frequently chosen close to
$\boldsymbol{\mu}^{\ast}$ so that there is some constant $\gamma_{1}>0$ with
\[
\lim_{N\rightarrow+\infty}\frac{1}{N}\sum_{n=1}^{N}\mathbb{I}\left\{
\left\vert \bar{\mu}_{i}\left(  \mathbf{q}(n)\right)  -\mu_{i}^{\ast
}\right\vert <\theta,\forall i\in\mathcal{M}\right\}  \geq\gamma_{1}.
\]
for arbitrary $\theta > 0 $ and any $\mathbf{q}(n)$, $n \in
\mathbb{N}$, with probability $1$.

If $\left\vert \bar{\mu}_{i}\left(  \mathbf{q}(n)\right)  -\mu_{i}^{\ast
}\right\vert <\theta$, according to the definition of $\bar{\boldsymbol{\mu}%
}\left(  \mathbf{q}\right)  $ we have
\begin{equation}
C_{1}\leq\frac{\mu_{i}^{\ast}-\theta}{\mu_{j}^{\ast}+\theta}\leq\frac{\bar
{\mu}_{i}\left(  \mathbf{q}(n)\right)  }{\bar{\mu}_{j}\left(  \mathbf{q}%
(n)\right)  }=\frac{e^{q_{i}}}{e^{q_{j}}}\leq\frac{\mu_{i}^{\ast}+\theta}%
{\mu_{j}^{\ast}-\theta}\leq C_{2}, \label{eqn:exp_weight}%
\end{equation}
for some constant $C_{1},C_{2}>0$. We denote by
\[
\Delta\mathbf{q}:=\mathbf{q}(n+1)-\mathbf{q}(n)=\mathbf{a}(n)-\mathbf{r}%
^{\mathcal{P}}(\mathbf{h}(n),\mathbf{q}(n)).
\]
Due to the randomness of $\mathbf{a}(n)$ and $\mathbf{h}(n)$, for a particular
user $i\in\mathcal{M}$, there is some probability $\gamma_{2}>0$\ that
\begin{equation}
\Pr\left\{  C_{3}<\Delta q_{i}<C_{4},\Delta q_{j}\leq0,\forall
j\neq i\right\}  \geq\gamma_{2} \label{ineq:gamma2}%
\end{equation}
for some constants $C_{3},C_{4}>0$. Then it holds
\begin{align*}
&  \bar{\mu}_{i}\left(  \mathbf{q}(n+1)\right)  -\bar{\mu}_{i}\left(
\mathbf{q}(n)\right) \\
=  &  \frac{e^{q_{i}+\Delta q_{i}}}{e^{q_{i}+\Delta q_{i}}+\sum_{j\neq
i}e^{q_{j}+\Delta q_{j}}}-\frac{e^{q_{i}}}{e^{q_{i}}+\sum_{j\neq i}e^{q_{j}}%
}\\
\geq &  \frac{\left(  e^{q_{i}+\Delta q_{i}}-e^{q_{i}}\right)  \sum_{j\neq
i}e^{q_{j}}}{\left(  e^{q_{i}+\Delta q_{i}}+\sum_{j\neq i}e^{q_{j}}\right)
\left(  e^{q_{i}}+\sum_{j\neq i}e^{q_{j}}\right)  }\\
\geq &  \frac{\left(  e^{\Delta q_{i}}-1\right)  }{\left(  e^{\Delta q_{i}%
}+\sum_{j\neq i}\frac{e^{q_{j}}}{e^{qi}}\right)  \left(  \frac{e^{q_{i}}}%
{\sum_{j\neq i}e^{q_{j}}}+1\right)  }.
\end{align*}
According to (\ref{eqn:exp_weight}) we have
\begin{align*}
&  \bar{\mu}_{i}\left(  \mathbf{q}(n+1)\right)  -\bar{\mu}_{i}\left(
\mathbf{q}(n)\right) \\
\geq &  \frac{e^{C_{3}}-1}{\left(  e^{C_{4}}+\frac{M-1}{C_{1}}\right)  \left(
\frac{C_{2}}{M-1}+1\right)  }=\epsilon
\end{align*}
for some constant $\epsilon>0$. Then, combining with
(\ref{ineq:gamma2}) and our i.i.d. assumption the inequality
(\ref{condition_necessity_C1_compl}) in Theorem
\ref{theorem:necessity} holds
with%
\begin{align*}
&  \lim_{N\rightarrow+\infty}\frac{1}{N}\sum_{n=1}^{N}\mathbb{I}\left\{
\bar{\mu}_{i}\left(  \mathbf{q}(n+1)\right)  -\bar{\mu}_{i}\left(
\mathbf{q}(n)\right)  >\epsilon,\exists i\in\mathcal{M}\right\} \\
&  \geq\gamma_{1}\gamma_{2}%
\end{align*}
with probability $1$ and the queueing system is not stable.

\section{Conclusions}

\label{section:conclusions}

We have presented sufficient and necessary conditions for
throughput-optimality of queue-length based scheduling policies. For a wide
class of arrival and channel models these conditions guarantee that if the
mean arrival rate lies inside the ergodic achievable rate region, the system
is stable. With application examples such as Queue Proportional Scheduling and
Idle State Prediction Scheduling we have shown that the stability can even
been proven in cases where conventional proof techniques fail.

\section{Appendix}

\subsection{Proof of Theorem \ref{theorem:convergence}}

Stability can be proven by checking the so-called \emph{Lyapunov drift
criteria} as given in \cite{Tassiulas:1992,NeeModRoh:2003,Leonardi:2001}. That
is to say if we can find some non-negative $V(\mathbf{q}):\mathbb{R}_{+}%
^{M}\rightarrow\mathbb{R}_{+}$, some $\theta>0$ and a compact region
$\bar{\mathcal{B}}:=\left\{  \mathbf{q}:\Vert\mathbf{q}\Vert\leq B\right\}  $
such that
\begin{align}
\quad &  \mathbb{E}\left\{  \left.  V\left(  \mathbf{q}\left(  n+1\right)
\right)  \right\vert \mathbf{q}\left(  n\right)  \right\}  < +\infty
\qquad\quad & \forall\mathbf{q}\left(  n\right)  \in\bar{\mathcal{B}%
}\label{condition_Lyapunov_1}\\
&  \Delta V\left(  \mathbf{q}\left(  n\right)  \right)  < - \theta &
\forall\mathbf{q}\left(  n\right)  \notin\bar{\mathcal{B}},
\label{condition_Lyapunov_2}%
\end{align}
the queueing system is positive recurrent. Here, $\Delta V\left(
\mathbf{q}\left(  n\right)  \right)  $ is the one-step drift defined as
\[
\Delta V\left(  \mathbf{q}\left(  n\right)  \right)  :=\mathbb{E}\left\{
\left.  V\left(  \mathbf{q}\left(  n+1\right)  \right)  -V\left(
\mathbf{q}\left(  n\right)  \right)  \right\vert \mathbf{q}\left(  n\right)
\right\}  .
\]
Furthermore, if for some $\theta>0$, it satisfies
\begin{equation}
\Delta V\left(  \mathbf{q}\right)  \leq-\theta f\left(  \mathbf{q}\right)
,\qquad\forall\Vert\mathbf{q}\Vert>B \label{ineq:negative_drift}%
\end{equation}
for some $B>0$ and unbounded positive function $f\left(  \mathbf{q}\right)  $,
it can be shown that the queueing system is f-stable.

We carry out the proof in two steps. First, we prove the throughput-optimality
for those policies, whose weight factors $\bar{\boldsymbol{\mu}}(\mathbf{q})$
fulfill the integrability condition in Eqn.(\ref{condition_integrable}). The
weight factors in those policies can be regarded as the normalized gradient of
a certain potential field $V(\mathbf{q})$. We show that the expected drift
$\Delta V\left(  \mathbf{q}\right)  $ satisfies the inequality
(\ref{ineq:negative_drift}) and hence the system driven by those policies is
stable. In the second step, we extend the results to all other policies whose
weight factors are not integrable. It is shown that if the policies fulfill
the condition given in the theorem, their weight factors $\bar{\boldsymbol{\mu
}}(\mathbf{q})$ can be approximated by some functions $\tilde{\boldsymbol{\mu
}}(\mathbf{q})$ which are integrable. Then we prove the drift condition
$\Delta V\left(  \mathbf{q}\right)  $ for those policies and establish the stability.

Firstly, we analyze the subclass of weight functions whose $\bar{\mu}%
_{i}(\mathbf{q})$ are continuously differentiable. Furthermore, we assume that
the weight functions satisfy the integrability condition, i.e.,
\begin{equation}
\frac{\partial\left(  \bar{\mu}_{i}\left(  \mathbf{q}\right)  \right)
}{\partial q_{j}} = \frac{\partial\left(  \bar{\mu}_{j}\left(  \mathbf{q}%
\right)  \right)  }{\partial q_{i}}, \qquad\forall i,j \in\mathcal{M}.
\label{condition_integrable}%
\end{equation}
For scheduling policies with this kind of weight functions, we have the
following lemma.

\begin{lemma}
If Eqn. (\ref{condition_integrable}) holds for all $\mathbf{q} \in
\mathbb{R}^{M}_{+}$ , then any vector of arrival rates $\boldsymbol{\rho}%
\in\text{int} (\mathcal{C}( \hat{P}))$ is stabilizable under the corresponding
scheduling policy as long as $\bar{\boldsymbol{\mu} }(\mathbf{q})$ fulfills
the conditions given in Theorem \ref{theorem:convergence}.
\label{lemma_stable_integrable}
\end{lemma}

\begin{proof}
Condition (\ref{condition_integrable}) implies that the vector field defined
by $\bar{\boldsymbol{\mu}}(\mathbf{q})$ has the path independence property,
namely the integral of $\bar{\boldsymbol{\mu}}(\mathbf{q})$ along a path
depends only on the start and end points of that path, not the particular
route taken. According to \emph{Poincar{\'e} lemma}, the vector field
$\bar{\boldsymbol{\mu}}(\mathbf{q})$ is completely integrable and it is the
gradient of a scalar field, that is to say, there exist some function
$f(\mathbf{q}):\mathbb{R}_{+}^{M}\rightarrow\mathbb{R}_{+}$ with
\begin{equation}
\frac{\partial f(\mathbf{q})}{\partial q_{i}}=\bar{\mu}_{i}(\mathbf{q}).
\end{equation}
Setting the value of $f(\mathbf{q})$ at the origin equal zero, $f(\mathbf{q})$
at the point $\mathbf{q}$ can be calculated by
\begin{equation}
f(\mathbf{q})=\int_{0}^{\Vert\mathbf{q}\Vert_{2}}\bar{\boldsymbol{\mu}}\left(
t\bar{\mathbf{q}}\right)  ^{T}\bar{\mathbf{q}}dt, \label{eqn:f_integral}%
\end{equation}
where $\bar{\mathbf{q}}:=\frac{\mathbf{q}}{\Vert\mathbf{q}\Vert_{2}}$ is the
normalized vector of $\mathbf{q}$. Since each element of $\bar{\boldsymbol{\mu
}}(\mathbf{q})$ is larger than or equal to zero, $f(\mathbf{q})$ is a positive
function. Moreover, if $\|\mathbf{q}\|$ becomes large, according to Condition
2) in the Theorem \ref{theorem:convergence}, for $i$-th queue with bounded
$q_{i}$, $\bar{q}_{i} \rightarrow0$ results in $\bar{\mu}_{i}(\mathbf{q})
\rightarrow0$. Then for other queues with $\bar{\mu}_{j}(\mathbf{q}) > C_{\mu
}$, $q_{j}$ grows with increasing $\|\mathbf{q}\|$ and we have $\bar{q}_{j} >
C_{q}$ for some $C_{q}$ and $C_{\mu} > 0$. Thus it holds
\[
\bar{\boldsymbol{\mu}}(\mathbf{q})^{T} \mathbf{\bar{q}} > C
\]
for some $C>0$ if $\|\mathbf{q}\|$ is sufficiently large. Considering
Eqn.(\ref{eqn:f_integral}), it follows that $f(\mathbf{q})\rightarrow+\infty$
as $\left\Vert \mathbf{q}\right\Vert \rightarrow+\infty$. Therefore,
$f(\mathbf{q})$ is a positive, unbounded function as we used in Definition
\ref{definition_f_stable}.

Observing a new vector field defined by $\boldsymbol{\nu}(\mathbf{q}%
)=f(\mathbf{q})\bar{\boldsymbol{\mu}}(\mathbf{q})$, we have
\begin{align}
\frac{\partial\left(  \nu_{i}\left(  \mathbf{q}\right)  \right)  }{\partial
q_{j}}  &  =\frac{\partial\left(  f\left(  \mathbf{q}\right)  \bar{\mu}%
_{i}\left(  \mathbf{q}\right)  \right)  }{\partial q_{j}}\nonumber\\
&  =\bar{\mu}_{j}\left(  \mathbf{q}\right)  \bar{\mu}_{i}\left(
\mathbf{q}\right)  +\frac{\partial\bar{\mu}_{j}\left(  \mathbf{q}\right)
}{\partial q_{i}}\nonumber\\
&  =\bar{\mu}_{j}\left(  \mathbf{q}\right)  \bar{\mu}_{i}\left(
\mathbf{q}\right)  +\frac{\partial\bar{\mu}_{i}\left(  \mathbf{q}\right)
}{\partial q_{j}}\nonumber\\
&  =\frac{\partial\left(  f\left(  \mathbf{q}\right)  \bar{\mu}_{j}\left(
\mathbf{q}\right)  \right)  }{\partial q_{i}}\nonumber\\
&  =\frac{\partial\left(  \nu_{j}\left(  \mathbf{q}\right)  \right)
}{\partial q_{i}},\qquad\forall i,j\in\mathcal{M}.
\label{condition_integrable2}%
\end{align}
Condition (\ref{condition_integrable2}) ensures that $\boldsymbol{\nu
}(\mathbf{q})$ is also the gradient of a scalar field and there is a function
$V(\mathbf{q}):\mathbb{R}_{+}^{M}\rightarrow\mathbb{R}_{+}$ with
\[
\frac{\partial V(\mathbf{q})}{\partial q_{i}}=f(\mathbf{q})\bar{\mu}%
_{i}(\mathbf{q}),
\]
where $f(\mathbf{q})$ is the magnitude of the gradient and $\boldsymbol{\bar
{\mu}}(\mathbf{q})$ is the direction of the gradient. Set
$V(\mathbf{{\boldsymbol{0}}}) = 0$ and $V(\mathbf{q})$ at the point
$\mathbf{q}$ is
\[
V(\mathbf{q})=\int_{0}^{\Vert\mathbf{q}\Vert_{2}}f\left(  t\bar{\mathbf{q}%
}\right)  \bar{\boldsymbol{\mu}}\left(  t\bar{\mathbf{q}}\right)  ^{T}%
\bar{\mathbf{q}}dt.
\]
It is easy to shown that the function $V(\mathbf{q})$ is also a positive,
unbounded function. We use the function $V(\mathbf{q})$ as our Lyapunov
function in the proof.

The first condition of the Lyapunov function given in
(\ref{condition_Lyapunov_1}) is satisfied as long as the arrival rates
$a_{i}(n)$ and transmission rates $r_{i}(n)$ are bounded. Next we analyze the
second condition, namely the drift of $V(\mathbf{q})$ of the queueing system.
For convenience we use the superscript to denote the time slot in the following.

Using the\textit{ mean value theorem} of differential calculus we have for
some $\mathbf{\tilde{q}}^{n}$ between $\mathbf{q}^{n}$ and $\mathbf{q}^{n+1}$
i.e. $\tilde{q}_{i}^{n}=\alpha_{i}q_{i}^{n}+\left(  1-\alpha_{i}\right)
q_{i}^{n+1}$, $\forall i\in\mathcal{M}$, for some $\alpha_{i}\in\lbrack0,1]$
\begin{align}
&  \Delta V(\mathbf{q}^{n})\\
=  &  \mathbb{E}\left\{  \left.  \sum_{i=1}^{M}f(\tilde{\mathbf{q}}^{n}%
)\bar{\mu}_{i}(\tilde{\mathbf{q}}^{n})\left(  a_{i}^{n}-r_{i}^{n}\right)
\right\vert \mathbf{q}^{n}\right\} \nonumber\\
&  +\mathbb{E}\left\{  \left.  \sum_{i=1}^{M}f(\tilde{\mathbf{q}}^{n})\bar
{\mu}_{i}(\tilde{\mathbf{q}}^{n})z_{i}^{n}\right\vert \mathbf{q}^{n}\right\}
\label{term:deltav}%
\end{align}

Considering the first part in (\ref{term:deltav}), we have
\begin{align}
&  \mathbb{E}\left\{  \left.  \sum_{i=1}^{M}f(\tilde{\mathbf{q}}^{n})\bar{\mu
}_{i}(\tilde{\mathbf{q}}^{n})\left(  a_{i}^{n}-r_{i}^{n}\right)  \right\vert
\mathbf{q}^{n}\right\} \\
\leq &  f(\mathbf{q}^{n})\left(  \sum_{i=1}^{M}\bar{\mu}_{i}(\mathbf{q}%
^{n})\rho_{i}-\sum_{i=1}^{M}\bar{\mu}_{i}(\mathbf{q}^{n})\mathbb{E}\left\{
\left.  r_{i}^{n}\right\vert \mathbf{q}^{n}\right\}  \right) \nonumber\\
&  +\mathbb{E}\left\{  \left.  \sum_{i=1}^{M}|f(\tilde{\mathbf{q}}^{n}%
)\bar{\mu}_{i}(\tilde{\mathbf{q}}^{n})-f(\mathbf{q}^{n})\bar{\mu}%
_{i}(\mathbf{q}^{n})||a_{i}^{n}-r_{i}^{n}|\right\vert \mathbf{q}^{n}\right\}
. \label{term:sub1}%
\end{align}
Since
\[
\mathbb{E}\left\{  \left.  \mathbf{r}^{n}\right\vert \mathbf{q}^{n}\right\}
=\arg\max_{\widetilde{\mathbf{r}}\in\mathcal{C}\left(  \hat{P}\right)
}\boldsymbol{\bar{\mu}}(\mathbf{q}^{n})^{T}\widetilde{\mathbf{r}}.
\]
for any $\boldsymbol{\rho}\in\text{int} (\mathcal{C}\left(  \hat{P}\right) )
$, we can always find some $\Gamma>0$, so that
\[
\mathbb{E}\left\{  \left.  \sum_{i=1}^{M}\bar{\mu}_{i}(\mathbf{q})(\rho
_{i}-r_{i}^{n})\right\vert \mathbf{q}^{n}\right\}  \leq-\Gamma.
\]
Hence the first part in (\ref{term:sub1})
\begin{align*}
&  f(\mathbf{q}^{n})\left(  \sum_{i=1}^{M}\bar{\mu}_{i}(\mathbf{q}^{n}%
)\rho_{i}-\sum_{i=1}^{M}\bar{\mu}_{i}(\mathbf{q}^{n})\mathbb{E}\left\{
\left.  r_{i}^{n}\right\vert \mathbf{q}^{n}\right\}  \right) \\
\leq &  -\Gamma f(\mathbf{q}^{n}).
\end{align*}

For the second part in (\ref{term:sub1}), we define $\Delta\mathbf{q}%
=\tilde{\mathbf{q}}^{n}-\mathbf{q}^{n}$. Then
\begin{align*}
&  f(\mathbf{q}^{n}+\Delta\mathbf{q})-f(\mathbf{q}^{n})\\
=  &  \int_{0}^{1}\boldsymbol{\bar{\mu}}(\mathbf{q}^{n}+t\Delta\mathbf{q}%
)\Delta\mathbf{q}dt\\
\leq &  \int_{0}^{1}\left\Vert \Delta\mathbf{q}\right\Vert _{1}dt=\left\Vert
\Delta\mathbf{q}\right\Vert _{1}%
\end{align*}
Since $a_{i}^{n}$ and $r_{i}^{n}$ are bounded, we choose some $C_{3}>1$ so
that $a_{i}^{n}<C_{3}$ and $r_{i}^{n}<C_{3}$ for all $i$. Then $\left\Vert
\Delta\mathbf{q}\right\Vert _{1}$ is bounded by $2MC_{3}$ and we have
\[
\left\vert f\left(  \tilde{\mathbf{q}}^{n}\right)  -f\left(  \mathbf{q}%
^{n}\right)  \right\vert <\epsilon_{3}f\left(  \mathbf{q}\right)
\]
for any given $\epsilon_{3}>0$ and sufficiently large $\Vert\mathbf{q}\Vert$.
According to Condition 1) in Theorem \ref{theorem:convergence}, we also have
\[
\left\vert \bar{\mu}_{i}(\tilde{\mathbf{q}}^{n})-\bar{\mu}_{i}(\mathbf{q}%
^{n})\right\vert <\epsilon_{1}.
\]
Then if $\Vert\mathbf{q}^{n}\Vert$ is sufficiently large,
\begin{align}
&  \mathbb{E}\left\{  \left.  \sum_{i=1}^{M}|f(\tilde{\mathbf{q}}^{n})\bar
{\mu}_{i}(\tilde{\mathbf{q}}^{n})-f(\mathbf{q}^{n})\bar{\mu}_{i}%
(\mathbf{q}^{n})||a_{i}^{n}-r_{i}^{n}|\right\vert \mathbf{q}^{n}\right\}
\nonumber\\
\leq &  2C_{3}\mathbb{E}\left\{  \left.  \sum_{i=1}^{M}\left(  f(\mathbf{q}%
^{n})+\epsilon_{3}f(\mathbf{q}^{n})\right)  \left(  \bar{\mu}_{i}%
(\mathbf{q}^{n})+\epsilon_{1}\right)  \right\vert \mathbf{q}^{n}\right\}
\nonumber\\
&  -2C_{3}\mathbb{E}\left\{  \left.  \sum_{i=1}^{M}f(\mathbf{q}^{n})\bar{\mu
}_{i}(\mathbf{q}^{n})\right\vert \mathbf{q}^{n}\right\} \nonumber\\
=  &  \underset{\sigma_{1}}{\underbrace{\left(  2MC_{3}\epsilon_{1}%
+2C_{3}\epsilon_{3}+2MC_{3}\epsilon_{1}\epsilon_{3}\right)  }}f(\mathbf{q}%
^{n}) \label{term:sub12}%
\end{align}
holds for any $\epsilon_{1}$, $\epsilon_{3}>0$. Hence we have $\sigma
_{1}\rightarrow0$ when $\Vert\mathbf{q}\Vert\rightarrow+\infty$.

Now we consider the second part in (\ref{term:deltav}).
\begin{align}
&  \mathbb{E}\left\{  \left.  \sum_{i=1}^{M}f(\tilde{\mathbf{q}}^{n})\bar{\mu
}_{i}(\tilde{\mathbf{q}}^{n})z_{i}^{n}\right\vert \mathbf{q}^{n}\right\} \\
\leq &  \mathbb{E}\left\{  \left.  \sum_{i=1}^{M}f(\mathbf{q}^{n})\bar{\mu
}_{i}(\mathbf{q}^{n})z_{i}^{n}\right\vert \mathbf{q}^{n}\right\} \nonumber\\
&  +\mathbb{E}\left\{  \left.  \sum_{i=1}^{M}\left\vert f(\tilde{\mathbf{q}%
}^{n})\bar{\mu}_{i}(\tilde{\mathbf{q}}^{n})-f(\mathbf{q}^{n})\bar{\mu}%
_{i}(\mathbf{q}^{n})\right\vert z_{i}^{n}\right\vert \mathbf{q}^{n}\right\}
\label{term:sub3}%
\end{align}

For the first part in (\ref{term:sub3}), since $z_{i}^{n}\leq r_{i}^{n}$ is
bounded by the current rate region, $\mathbb{E}\left\{  \left.  z_{i}%
^{n}\right\vert \mathbf{q}^{n}\right\}  $ is bounded by the ergodic achievable
rate region so that for some $C_{4}>0$ we have
\begin{equation}
\mathbb{E}\left\{  z_{i}(t)\right\}  \leq C_{4}.
\end{equation}
We define the set $\mathcal{G}:=\{i:z_{i}>0,i\in\mathcal{M}\}$. Since
$r_{i}^{n}<C_{3}$ is bounded by $C_{3}$, then $q_{i}^{n}<C_{3}$, $\forall
i\in\mathcal{G}$. If $\left\Vert \mathbf{q}^{n}\right\Vert $ is sufficiently
large so that $\left\Vert \mathbf{q}^{n}\right\Vert >MC_{3}$, we can exclude
the case $\mathcal{G}=\mathcal{M}$. According to Condition 2) we have
$\bar{\mu}_{i}(\mathbf{q}^{n})\leq\epsilon_{2}$, $\forall i\in\mathcal{G}$ for
arbitrarily small $\epsilon_{2}$. Then
\begin{equation}
\mathbb{E}\left\{  \left.  \sum_{i\in\mathcal{G}}f(\mathbf{q}^{n})\bar{\mu
}_{i}(\mathbf{q}^{n})z_{i}^{n}\right\vert \mathbf{q}^{n}\right\}
<MC_{4}\epsilon_{2}f(\mathbf{q}^{n})
\end{equation}
holds.

Using the same proof method as for (\ref{term:sub12}) it can be shown that the
second part in (\ref{term:sub3}) can be bounded by $\sigma_{2}f(\mathbf{q}%
^{n})$ for any $\sigma_{2}>0$.

Define $\theta=\Gamma-\sigma_{1}-MC_{4}\epsilon_{2}-\sigma_{2}$ and choose
$\sigma_{1}$, $\sigma_{2}$, $\epsilon_{2}$ so that $\theta>0$ we have the
drift
\begin{equation}
\Delta V(\mathbf{q}^{n})\leq-\theta f(\mathbf{q}^{n}) \label{ineq:Lyapunov}%
\end{equation}
and which is negative and the Markov chain is positive recurrent.
\end{proof}

Lemma \ref{lemma_stable_integrable} is applied to weight functions which are
completely integrable. In general the weight functions don't have to meet the
integrability condition (\ref{condition_integrable}) or can be even not
continuously differentiable. However, it can be shown that if the weight
function $\bar{\boldsymbol{\mu}}(\mathbf{q})$ has the properties described in
Theorem \ref{theorem:convergence}, it can be approximated by some (at least
piecewise integrable) function $\tilde{\boldsymbol{\mu}}(\mathbf{q})$. The
following lemma help us to achieve our main result.

\begin{lemma}
If the function $\bar{\boldsymbol{\mu}}(\mathbf{q})$ fulfills the Condition
1), 2) in Theorem \ref{theorem:convergence}, then there exists a positive,
unbounded function $f:\mathbb{R}^{M}_{+} \rightarrow\mathbb{R}_{+}$ as given
in Definition \ref{definition_f_stable}, and a positive, continuous, piecewise
differentiable function $V:\mathbb{R}^{M}_{+} \rightarrow\mathbb{R}_{+}$, such
that it holds
\begin{equation}
\frac{\partial V(\mathbf{q})}{\partial q_{i}}=f(\mathbf{q})\widetilde{\mu}%
_{i}(\mathbf{q}),\forall i\in\mathcal{M}%
\end{equation}
on each differentiable subdomain of $V$, and
\begin{equation}
\left\vert \tilde{\mu}_{i}(\mathbf{q})-\bar{\mu}_{i}(\mathbf{q})\right\vert
<\epsilon_{4},\forall i\in\mathcal{M},
\end{equation}
for any $\epsilon_{4}>0$ if $\Vert\mathbf{q}\Vert$ is sufficiently large.
\label{lemma_approx_mu}
\end{lemma}

\begin{proof}
In the following we show how to construct the function $V(\mathbf{q})$,
$f(\mathbf{q})$ and $\tilde{\boldsymbol{\mu}}(\mathbf{q})$ based on
$\bar{\boldsymbol{\mu}}(\mathbf{q})$. Since we only need to ensure that
$|\tilde{\mu}_{i}(\mathbf{q})-\bar{\mu}_{i}(\mathbf{q})|<\epsilon_{4}$ for
large $\|\mathbf{q}\|$, it is sufficient to construct the functions on the
domain where $\|\mathbf{q}\| \geq B$ for sufficiently large $B$. The function
$V$ and $f$ on the domain $\|\mathbf{q}\| \leq B$ can be defined as any
positive, bounded, continuously differentiable function, which is continuous
on the boundary $\|\mathbf{q}\| = B$.

\begin{figure}[h]
\begin{center}
\psfrag{V0}{\tiny$\mathbf{q}^0$} \psfrag{V1}{\tiny$\mathbf{q}^*$}
\psfrag{qi}{\footnotesize$q_i$} \psfrag{qj}{\footnotesize$q_j$}
\psfrag{qk}{\footnotesize$q_k$}
\psfrag{qa}{\footnotesize$\mathbf{q}^a$}
\psfrag{qb}{\footnotesize$\mathbf{q}^b$}
\psfrag{qc}{\footnotesize$\mathbf{q}^c$}
\psfrag{qd}{\footnotesize$\mathbf{q}^d$}
\psfrag{qe}{\tiny$\mathbf{q}^e$} \psfrag{qf}{\tiny$\mathbf{q}^f$}
\psfrag{dlqi}{\tiny$\Delta Q_i$} \psfrag{dlqj}{\tiny$\Delta Q_j$}
\psfrag{dsqi}{\tiny$\Delta q_i$} \psfrag{dsqj}{\tiny$\Delta q_j$}
\includegraphics[width=1\linewidth]{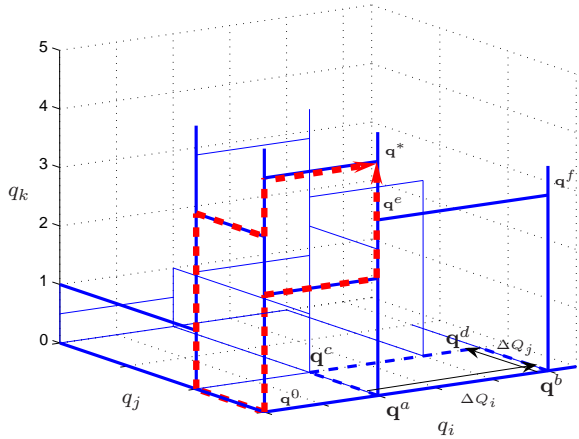}
\end{center}
\caption{Orthogonal grid (irregular) in $M=3$-dimension. The line integral
between two points on the grid line (e.g. along the two paths marked by dashed
line) depends only on the start point $\mathbf{q}^{0}$ and end point
$\mathbf{q}^{*}$. It is independent of the chosen pathes}%
\label{fig_grid_3D}%
\end{figure}

In the domain of $\Vert\mathbf{q}\Vert\geq B$, we at first construct a
orthogonal grid such that each cell in the grid is a rectangle (see
Fig.\ref{fig_grid_3D} for an example in $M=3$-dimension). Start by a point
$\mathbf{q}^{a}=\mathbf{Q}\in\mathbb{R}_{+}^{M}$, the next cell in the
dimensions $i$, $j$ (see Fig.\ref{fig_grid_2D}) has the grid points
\begin{align*}
\mathbf{q}^{a}  &  =[Q_{1},...,Q_{i},...,Q_{j},...,Q_{M}]^{T},\\
\mathbf{q}^{b}  &  =[Q_{1},...,Q_{i}+\Delta Q_{i},...,Q_{j},...,Q_{M}]^{T},\\
\mathbf{q}^{c}  &  =[Q_{1},...,Q_{i},...,Q_{j} + \Delta Q_{j},...,Q_{M}%
]^{T},\\
\mathbf{q}^{d}  &  =[Q_{1},...,Q_{i}+\Delta Q_{i},...,Q_{j} + \Delta
Q_{j},...,Q_{M}]^{T}.
\end{align*}
The length of the cell $\Delta Q_{i},\Delta Q_{j}$ is determined by the
equation
\begin{align}
&  \int_{0}^{\Delta Q_{i}}\bar{\mu}_{i}(...,q_{i},Q_{j},...)-\bar{\mu}%
_{i}(...,q_{i},Q_{j}+\Delta Q_{j},...)dq_{i}\nonumber\\
=  &  \int_{0}^{\Delta Q_{j}}\bar{\mu}_{j}(...,Q_{i},q_{j},...)-\bar{\mu}%
_{j}(...,Q_{i}+\Delta Q_{i},q_{j},...)dq_{j}.
\label{condition_path_independent}%
\end{align}

Condition 2) in Theorem \ref{theorem:convergence} implies that in the region
$\|\mathbf{q}\| \geq B$ for some large constant $B$, the function $\bar{\mu
}_{i}(\mathbf{q})$ decreases with increasing $q_{j}$ and $\bar{\mu}%
_{j}(\mathbf{q})$ decreases with increasing $q_{i}$ as well. Hence
\begin{align*}
&  \bar{\mu}_{i}(...,q_{i},Q_{j},...)-\bar{\mu}_{i}(...,q_{i},Q_{j}+\Delta
Q_{j},...) >0\\
&  \bar{\mu}_{j}(...,Q_{i},q_{j},...)-\bar{\mu}_{j}(...,Q_{i}+\Delta
Q_{i},q_{j},...) >0
\end{align*}
and Eqn.(\ref{condition_path_independent}) has positive general
solutions with $\Delta Q_{i}$, $\Delta Q_{j} > 0$. Iteratively
take $q^{b}$, $q^{c}$ and $q^{d}$ as start point, we can extend
the grid until it covers the subdomain in the dimensions $i$, $j$.
Based on the existing grid lines in the dimensions $i$, $j$ (e.g.
the line $\bar{\mathbf{q}^a\mathbf{q}^b}$ in Fig.
\ref{fig_grid_3D}), we can repeat the process in a further
dimension $k$ and construct the grid in this dimension (the grid
$\mathbf{q}^a$-$\mathbf{q}^b$-$\mathbf{q}^e$-$\mathbf{q}^f$).
Since relationship of $\Delta Q_{i}$ and $\Delta Q_{j}$ is
determined by the definition of $\bar {\mu}_{i}(\mathbf{q})$ on
the particular points, each rectangle in the grid has different
height and width so that the constructed grid has a irregular
pattern.

Denote the path starts at $q^{a}$ via $q^{b}$ to $q^{d}$ as $S_{abd}$ and the
path starts at $q^{a}$ via $q^{c}$ to $q^{d}$ as $S_{abd}$,
Eqn.(\ref{condition_path_independent}) ensures that the integral of the
function $\bar{\boldsymbol{\mu}}(\mathbf{q})$ along the path $S_{abd}$ equals
the integral along the path $S_{acd}$, which is
\begin{align}
&  \int_{S_{abd}}\bar{\boldsymbol{\mu}}(\mathbf{q})\cdot d\mathbf{s}%
\nonumber\\
=  &  \int_{0}^{\Delta Q_{i}}\bar{\mu}_{i}(...,q_{i},Q_{j},...)dq_{i}%
\nonumber\\
&  +\int_{0}^{\Delta Q_{j}}\bar{\mu}_{j}(...,Q_{i}+\Delta Q_{i},q_{j}%
,...)dq_{j}\nonumber\\
=  &  \int_{0}^{\Delta Q_{j}}\bar{\mu}_{j}(...,Q_{i},q_{j},...)dq_{j}%
\nonumber\\
&  +\int_{0}^{\Delta Q_{i}}\bar{\mu}_{i}(...,q_{i},Q_{j}+\Delta Q_{j}%
,...)dq_{i}\nonumber\\
=  &  \int_{S_{acd}}\bar{\boldsymbol{\mu}}(\mathbf{q})\cdot d\mathbf{s.}
\label{condition_path_independent2}%
\end{align}
Since Eqn.(\ref{condition_path_independent2}) holds for all cells of the grid,
the integral between arbitrary two grid points along any grid line has the
same value. Hence the vector field $\bar{\boldsymbol{\mu}}(\mathbf{q})$ can be
considered as "path-independent" along the grid lines. Then we define a
function $f(\mathbf{q})$ whose value on the grid line as the integral of
$\bar{\boldsymbol{\mu}}(\mathbf{q})$ along the grid lines, i.e.
\[
f(\mathbf{q^{\ast}}):=f(\mathbf{Q}^{0})+\int_{S}\bar{\boldsymbol{\mu}%
}(\mathbf{q})\cdot d\mathbf{s},
\]
where $\mathbf{q^{\ast}}$ is a point on the grid line and $S$ is an arbitrary
path between $\mathbf{q^{\ast}}$ and the initial point $\mathbf{Q}^{0}$ along
the grid lines.

Define a new vector field by $\boldsymbol{\nu}(\mathbf{q}):=f(\mathbf{q}%
)\bar{\boldsymbol{\mu}}(\mathbf{q})$, the line integral of $\boldsymbol{\nu
}(\mathbf{q})$ along the path $S_{abc}$ is
\begin{align*}
\int_{S_{abd}}\boldsymbol{\nu}(\mathbf{q})\cdot d\mathbf{s}  &  \mathbf{=}%
\int_{S_{abd}}f(\mathbf{q})\bar{\boldsymbol{\mu}}(\mathbf{q})\cdot
d\mathbf{s}\\
&  =\int_{S_{abd}}f(\mathbf{q})df(\mathbf{q})\\
&  =\frac{1}{2}\left(  f^{2}\left(  \mathbf{q}^{d}\right)  -f^{2}\left(
\mathbf{q}^{a}\right)  \right) \\
&  =\int_{S_{acd}}\boldsymbol{\nu}(\mathbf{q})\cdot d\mathbf{s}.
\end{align*}
Thus the integral of the vector field $\boldsymbol{\nu}(\mathbf{q})$ between
two grid points along the grid lines is also independent of the chosen paths.
Then we define a scalar field $V(\mathbf{q})$ whose value on the grid line is
given by
\[
V(\mathbf{q^{\ast}}):=V(\mathbf{Q}^{0})+\int_{S}f(\mathbf{q})\bar
{\boldsymbol{\mu}}(\mathbf{q})\cdot d\mathbf{s}.
\]
The value of $f(\mathbf{Q}^{0})$ and $V(\mathbf{Q}^{0})$ at the initial point
$\mathbf{Q}^{0}$ can be choose as an arbitrary positive constant. Since
$\bar{\boldsymbol{\mu}}_{i}(\mathbf{q})\geq0,$ $\forall i\in\mathcal{M}$, we
have $f(\mathbf{q^{\ast}})\rightarrow+\infty$ and $V(\mathbf{q^{\ast}%
})\rightarrow+\infty$ as $\left\Vert \mathbf{q^{\ast}}\right\Vert
\rightarrow+\infty$.

Once the value of $V(\mathbf{q^{\ast}})$ is fixed on the grid lines, we obtain
the value of $V$ inside a grid cell by the linear interpolation of
$V(\mathbf{q^{\ast}})$ along the lines parallel to the diagonal line (see
Fig.\ref{fig_grid_2D}), i.e. in the lower triangle with $\frac{\Delta q_{i}%
}{\Delta Q_{i}}+\frac{\Delta q_{j}}{\Delta Q_{j}}<1$, $V$ is defined as
\begin{equation}
V(...,Q_{i}+\Delta q_{i},Q_{j}+\Delta q_{j},...)= K_{i}V(\mathbf{q}^{I})
+K_{j}V(\mathbf{q}^{J}), \label{eqn:V_cell_1}%
\end{equation}
where
\begin{align*}
K_{i}  &  =\frac{\Delta Q_{j}\Delta q_{i}}{\Delta Q_{j}\Delta q_{i}+\Delta
Q_{i}\Delta q_{j}},\\
K_{j}  &  =\frac{\Delta Q_{i}\Delta q_{j}}{\Delta Q_{j}\Delta q_{i}+\Delta
Q_{i}\Delta q_{j}},\\
\mathbf{q}^{I}  &  =[Q_{1},...,Q_{i}+\Delta q_{i}+\frac{\Delta Q_{i}}{\Delta
Q_{j}}\Delta q_{j},Q_{j},...,Q_{M}]^{T},\\
\mathbf{q}^{J}  &  =[Q_{1},...,Q_{i},Q_{j}+\Delta q_{j}+\frac{\Delta Q_{j}%
}{\Delta Q_{i}}\Delta q_{i},...,Q_{M}]^{T}%
\end{align*}
and in the higher triangle with $\frac{\Delta q_{i}}{\Delta Q_{i}}%
+\frac{\Delta q_{j}}{\Delta Q_{j}}\geq1$, $V$ is defined as
\begin{equation}
V(...,Q_{i}+\Delta q_{i},Q_{j}+\Delta q_{j},...)=K_{i}V(\mathbf{q}^{I})
+K_{j}V(\mathbf{q}^{J}), \label{eqn:V_cell_2}%
\end{equation}
where%
\begin{align*}
K_{i}  &  =\frac{\Delta Q_{j}\Delta Q_{i}-\Delta Q_{j}\Delta q_{i}}{2\Delta
Q_{i}\Delta Q_{j}-\Delta Q_{j}\Delta q_{i}-\Delta Q_{i}\Delta q_{j}},\\
K_{j}  &  =\frac{\Delta Q_{j}\Delta Q_{i}-\Delta Q_{i}\Delta q_{j}}{2\Delta
Q_{i}\Delta Q_{j}-\Delta Q_{j}\Delta q_{i}-\Delta Q_{i}\Delta q_{j}},\\
\mathbf{q}^{I}  &  =[...,Q_{i}+\Delta q_{i}+\frac{\Delta Q_{i}}{\Delta Q_{j}%
}\Delta q_{j}-\Delta Q_{i},Q_{j}+\Delta Q_{j},...]^{T},\\
\mathbf{q}^{J}  &  =[...,Q_{i}+\Delta Q_{i},Q_{j}+\Delta q_{j}+\frac{\Delta
Q_{j}}{\Delta Q_{i}}\Delta q_{i}-\Delta Q_{j},...]^{T}.
\end{align*}

Eqn.(\ref{eqn:V_cell_1}) and (\ref{eqn:V_cell_2}) determine the value of
$V(\mathbf{q})$ on the orthogonal planes stretched by the grid, then the value
of $V(\mathbf{q})$ in the space between these planes is calculated by the
linear interpolation of the existing value in further dimensions. Similarly,
we can also define the value of $f(\mathbf{q})$ in the entire domain.

\begin{figure}[h]
\begin{center}
\psfrag{qi}{$q_i$} \psfrag{qj}{$q_j$} \psfrag{qa}{\footnotesize$\mathbf{q}^a$}
\psfrag{qb}{\footnotesize$\mathbf{q}^b$}
\psfrag{qc}{\footnotesize$\mathbf{q}^c$}
\psfrag{qd}{\footnotesize$\mathbf{q}^d$}
\psfrag{dlqi}{\footnotesize$\Delta Q_i$}
\psfrag{dlqj}{\footnotesize$\Delta Q_j$} \psfrag{dsqi}{\tiny$\Delta q_i$}
\psfrag{dsqj}{\tiny$\Delta q_j$} \psfrag{V1}{\tiny$V(\mathbf{q}^{I})$}
\psfrag{V2}{\tiny$V(\mathbf{q}^{J})$} \psfrag{V3}{\tiny$V(\mathbf{q})$}
\includegraphics[width=1\linewidth]{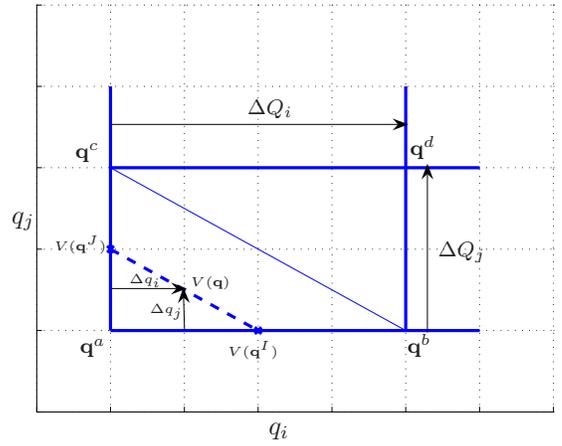}
\end{center}
\caption{The value of $V(\mathbf{q}):=V(...,Q_{i}+\Delta q_{i},Q_{j}+\Delta
q_{j},...)$ is calculated by the linear interpolation between the value
$V(\mathbf{q}^{I})$ and $V(\mathbf{q}^{I})$ defined on the grid lines. The
line $\overline{\mathbf{q}^{I}\mathbf{q}^{J}}$ is parallel to the diagonal
$\overline{\mathbf{q}^{b}\mathbf{q}^{c}}$ }%
\label{fig_grid_2D}%
\end{figure}

\begin{figure}[h]
\begin{center}
\psfrag{Vv}{$V(\mathbf{q})$} \psfrag{qi}{$q_i$} \psfrag{qj}{$q_j$}
\psfrag{V00}{\footnotesize$V(\mathbf{q}^a)$}
\psfrag{V10}{\footnotesize$V(\mathbf{q}^b)$}
\psfrag{V01}{\footnotesize$V(\mathbf{q}^c)$}
\psfrag{V11}{\footnotesize$V(\mathbf{q}^d)$}
\includegraphics[width=1\linewidth]{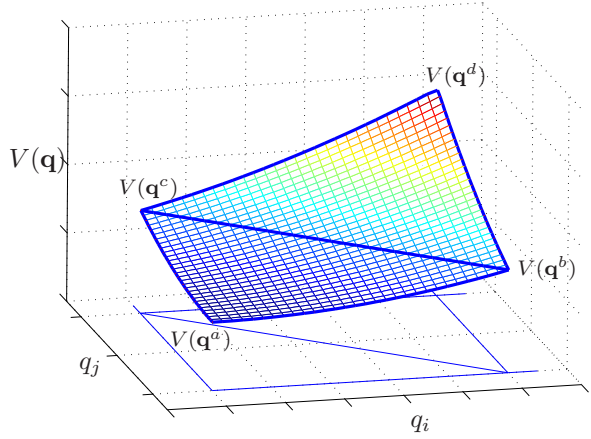}
\end{center}
\caption{The Lyapunov function $V(\mathbf{q})$ is differentiable inside the
subdomain between $\mathbf{q}^{a}$,$\mathbf{q}^{b}$,$\mathbf{q}^{c}$ and the
subdomain between $\mathbf{q}^{b}$,$\mathbf{q}^{c}$,$\mathbf{q}^{d}$}%
\label{fig_V_3D}%
\end{figure}

Observing the function $V(\mathbf{q})$, we can see that it is continuous in
$\mathbb{R}_{+}^{M}$ and differentiable in each subspace bounded by the grid
lines and diagonal lines. For two points $\mathbf{q}$ and $\mathbf{q}^{\prime
}$ which lie in the same cell, under Condition 1) in Theorem
\ref{theorem:convergence} we have $\left\vert \bar{\mu}_{i}\left(
\mathbf{q}\right)  -\bar{\mu}_{i}\left(  \mathbf{q}^{\prime}\right)
\right\vert \leq\epsilon_{1}$ and hence $\left\vert f\left(  \mathbf{q}%
\right)  -f\left(  \mathbf{q}^{\prime}\right)  \right\vert \leq\epsilon
_{1}f\left(  \mathbf{q}\right)  $ for arbitrarily small $\epsilon_{1}>0$. Then
for Eqn.(\ref{eqn:V_cell_1}) it holds
\begin{align*}
V(\mathbf{q}^{I})  &  =V(\mathbf{q}^{a})+f\left(  \mathbf{q}\right)  \left(
\bar{\mu}_{i}\left(  \mathbf{q}\right)  +\varepsilon_{i}(\mathbf{q})\right)
\left(  \Delta q_{i}+\frac{\Delta Q_{i}}{\Delta Q_{j}}\Delta q_{j}\right)  ,\\
V(\mathbf{q}^{J})  &  =V(\mathbf{q}^{a})+f\left(  \mathbf{q}\right)  \left(
\bar{\mu}_{j}\left(  \mathbf{q}\right)  +\varepsilon_{j}(\mathbf{q})\right)
\left(  \Delta q_{j}+\frac{\Delta Q_{j}}{\Delta Q_{i}}\Delta q_{i}\right)  .
\end{align*}
and further
\begin{align*}
V(\mathbf{q})=  &  V(\mathbf{q}^{a}) + f\left(  \mathbf{q}\right)  \left(
\bar{\mu}_{i}\left(  \mathbf{q}\right)  +\varepsilon_{i}(\mathbf{q})\right)
\Delta q_{i}\\
&  \qquad\quad+ f\left(  \mathbf{q}\right)  \left(  \bar{\mu}_{j}\left(
\mathbf{q}\right)  +\varepsilon_{j}(\mathbf{q})\right)  \Delta q_{j},
\end{align*}
where the deviation $\varepsilon_{i}(\mathbf{q}),\varepsilon_{j}%
(\mathbf{q})\rightarrow0$ as $\Vert\mathbf{q}\Vert\rightarrow+\infty$.
Similarly we can also obtain the same result for Eqn.(\ref{eqn:V_cell_2}).

Then the partial derivative of $V$ is
\[
\frac{\partial V(\mathbf{q})}{\partial q_{i}}=f\left(  \mathbf{q}\right)
\left(  \bar{\mu}_{i}\left(  \mathbf{q}\right)  +\epsilon_{4}\right)  .
\]
for arbitrarily small $\epsilon_{4}>0$ and we obtain the Lemma
\ref{lemma_approx_mu}.
\end{proof}

It can be shown that $f(\mathbf{q})$ and $V(\mathbf{q})$ constructed in Lemma
\ref{lemma_approx_mu} are positive and grow to infinity as $\left\Vert
\mathbf{q}\right\Vert \rightarrow+\infty$. Now we use the function
$V(\mathbf{q})$ and $f(\mathbf{q})$ in Lemma \ref{lemma_approx_mu} as the
Lyapunov function and the stability measure function respectively. It can also
be shown that $\Delta V(\mathbf{q^{n}})$ is bounded if $\mathbf{q^{n}}$ lies
in some compacted region $\bar{\mathcal{B}}$ and the arrival rates $a_{i}^{n}$
and transmission rates $r_{i}^{n}$ are bounded. Hence the Lyapunov condition
(\ref{condition_Lyapunov_1}) is satisfied.

\begin{figure}[h]
\begin{center}
\psfrag{qi}{$q_i$} \psfrag{qj}{$q_j$} \psfrag{q0}{\footnotesize$\mathbf{q}^n$}
\psfrag{q1}{\footnotesize$\mathbf{q}^{(1)}$}
\psfrag{q2}{\footnotesize$\mathbf{q}^{(2)}$}
\psfrag{q3}{\footnotesize$\mathbf{q}^{(3)}$}
\psfrag{q4}{\footnotesize$\mathbf{q}^{(4)}$}
\psfrag{q5}{\footnotesize$\mathbf{q}^{n+1}$}
\includegraphics[width=1\linewidth]{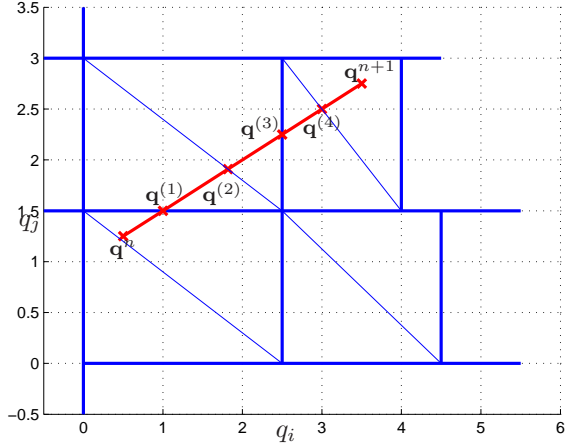}
\end{center}
\caption{The drift $\Delta V$ crosses 5 subdomains, which can be written as
the sum of the difference between $V(\mathbf{q}^{n})$, $V(\mathbf{q}^{(1)})$,
..., and $V(\mathbf{q}^{n+1})$}%
\label{fig_grid_2D_2}%
\end{figure}

Next we consider the drift $\Delta V(\mathbf{q}^{n})$ in Lyapunov condition
(\ref{condition_Lyapunov_2}) where $\mathbf{q^{n}} \notin\bar{\mathcal{B}}$.
The connection between $\mathbf{q}^{n}$ and $\mathbf{q}^{n+1}$ probably pass
through multiple differentiable subspaces of $V(\mathbf{q})$ (see
Fig.\ref{fig_grid_2D_2}), so we denote the intersection of the connecting line
and the boundary of the subspaces as $\mathbf{q}^{(1)},...,\mathbf{q}^{(L)}$
and the difference as $\Delta\mathbf{q}^{(1)}=\mathbf{q}^{(1)}-\mathbf{q}^{n}%
$,..., $\Delta\mathbf{q}^{(l)}=\mathbf{q}^{(1+1)}-\mathbf{q}^{(l)}$. The drift
is written as:%
\begin{align*}
&  \Delta V(\mathbf{q}^{n})\\
=  &  \mathbb{E}\left\{  V(\mathbf{q}^{n+1})-V(\mathbf{q}^{(L)})+\sum
_{l=2}^{L}V(\mathbf{q}^{l+1})-V(\mathbf{q}^{(l)})\right. \\
&  \quad+\Biggl.\left.  V(\mathbf{q}^{(1)})-V(\mathbf{q}^{n})\right\vert
\mathbf{q}^{n}\Biggr\}\\
=  &  \mathbb{E}\left\{  \left.  \sum_{l=1}^{L+1}f(\tilde{\mathbf{q}}%
^{(l)})\boldsymbol{\tilde{\mu}}(\tilde{\mathbf{q}}^{(l)})\cdot\Delta
\mathbf{q}^{(l)}\right\vert \mathbf{q}^{n}\right\} \\
\leq &  \mathbb{E}\left\{  \left.  \sum_{l=1}^{L+1}f(\tilde{\mathbf{q}}%
^{(l)})\bar{\boldsymbol{\mu}}(\tilde{\mathbf{q}}^{(l)})\cdot\Delta
\mathbf{q}^{(l)}+\epsilon_{4}\left\Vert \Delta\mathbf{q}^{(l)}\right\Vert
f(\tilde{\mathbf{q}}^{(l)})\right\vert \mathbf{q}^{n}\right\}  ,
\end{align*}
Where $\tilde{\mathbf{q}}^{(l)}$ is some point in the $l$-th subspace. Since
the arrival rates $a_{i}^{n}$ and the transmission rates $r_{i}^{n}$ are
bounded for all $i\in\mathcal{M}$, the difference $\Vert\Delta\mathbf{q}%
\Vert=\Vert\mathbf{q}^{n+1}-\mathbf{q}^{n}\Vert$ is bounded. Thus according to
Condition 1) in Theorem \ref{theorem:convergence} we have $\left\vert \bar
{\mu}_{i}(\tilde{\mathbf{q}}^{(l)})-\bar{\mu}_{i}(\tilde{\mathbf{q}}%
^{(1)})\right\vert <\epsilon_{1}$ and $\left\vert f(\tilde{\mathbf{q}}%
^{(l)})-f(\tilde{\mathbf{q}}^{(1)})\right\vert <\epsilon_{1}f\left(
\tilde{\mathbf{q}}^{(1)}\right)  $ for arbitrary $\epsilon_{1}>0$ if
$\Vert\tilde{\mathbf{q}}^{(1)}\Vert$ is large. The drift
\begin{align*}
&  \Delta V(\mathbf{q}^{n})\\
\leq &  \mathbb{E}\left\{  \left.  f(\tilde{\mathbf{q}}^{(1)})\bar
{\boldsymbol{\mu}}(\tilde{\mathbf{q}}^{(1)})\cdot\sum_{l=1}^{L+1}%
\Delta\mathbf{q}^{(l)}+\sigma_{3}f(\mathbf{q}^{n})\right\vert \mathbf{q}%
^{n}\right\} \\
\leq &  \mathbb{E}\left\{  \left.  f(\tilde{\mathbf{q}}^{(1)})\bar
{\boldsymbol{\mu}}(\tilde{\mathbf{q}}^{(1)})\cdot\left(  \mathbf{q}%
^{n+1}-\mathbf{q}^{n}\right)  \right\vert \mathbf{q}^{n}\right\}  +\sigma
_{3}f(\mathbf{q}^{n}),
\end{align*}
where $\sigma_{3}$ is some small constant.

Using the previous result in (\ref{ineq:Lyapunov}), it holds
\begin{align*}
&  \Delta V(\mathbf{q}^{n})\\
\leq &  \mathbb{E}\left\{  \left.  \sum_{i=1}^{M}f(\tilde{\mathbf{q}}^{n}%
)\bar{\mu}_{i}(\tilde{\mathbf{q}}^{n})\left(  a_{i}^{n}-r_{i}^{n}+z_{i}%
^{n}\right)  \right\vert \mathbf{q}^{n}\right\}  +\sigma_{3}f(\mathbf{q}%
^{n})\\
\leq &  -\theta f(\mathbf{q}^{n})+\sigma_{3}f(\mathbf{q}^{n})\\
\leq &  -\theta^{\prime}f(\mathbf{q}^{n})
\end{align*}
for some $\theta^{\prime}>0$ if $\left\Vert \mathbf{q}^{n}\right\Vert >B$, for
some $B>0$. The drift is negative thus the Markov chain is positive recurrent.

At last, we prove that the chain is also f-stable for the magnitude function
$f(\mathbf{q})$. We can write%
\begin{align*}
&  \mathbb{E}\left\{  \left.  V(\mathbf{q}^{n+1})\right\vert \mathbf{q}%
^{n}\right\} \\
\leq &  \mathbb{E}\left\{  \left.  V(\mathbf{q}^{n+1})\right\vert
\mathbf{q}^{n}>B\right\}  \Pr\left(  \mathbf{q}^{n}>B\right) \\
&  +\mathbb{E}\left\{  \left.  V(\mathbf{q}^{n+1})\right\vert \mathbf{q}%
^{n}\leq B\right\}  \Pr\left(  \mathbf{q}^{n}\leq B\right) \\
\leq &  \mathbb{E}\left\{  \left.  V(\mathbf{q}^{n})-\theta^{\prime
}f(\mathbf{q}^{n})\right\vert \mathbf{q}^{n}>B\right\}  \Pr\left(
\mathbf{q}^{n}>B\right) \\
&  +\mathbb{E}\left\{  \left.  V(\mathbf{q}^{n+1})\right\vert \mathbf{q}%
^{n}\leq B\right\}  \Pr\left(  \mathbf{q}^{n}\leq B\right) \\
\leq &  \mathbb{E}\left\{  V(\mathbf{q}^{n})\right\}  -\theta^{\prime
}f(\mathbf{q}^{n})+C_{5},
\end{align*}
where $C_{5}$ is some constant satisfying
\begin{align*}
C_{5}  &  \geq\mathbb{E}\left\{  \left.  V(\mathbf{q}^{n+1})\right\vert
\mathbf{q}^{n}\leq B\right\}  \Pr\left(  \mathbf{q}^{n}\leq B\right) \\
&  +\mathbb{E}\left\{  \left.  \theta^{\prime}f(\mathbf{q}^{n})\right\vert
\mathbf{q}^{n}\leq B\right\}  \Pr\left(  \mathbf{q}^{n}\leq B\right)  .
\end{align*}

Using the telescoping machinery, the summation of the drift over $N$ time
slots yields%
\[
\mathbb{E}\left\{  V(\mathbf{q}^{N})\right\}  \leq\mathbb{E}\left\{
V(\mathbf{q}^{1})\right\}  -\theta^{\prime}\sum_{n=1}^{N}\mathbb{E}\left\{
f(\mathbf{q}^{n})\right\}  +N\cdot C_{5}.
\]
since $V(\mathbf{q})$ is non-negative function, it holds%
\[
\sum_{n=1}^{N}\mathbb{E}\left\{  f(\mathbf{q}^{n})\right\}  \leq
\frac{\mathbb{E}\left\{  V(\mathbf{q}^{1})\right\}  }{\theta^{\prime}}%
+\frac{N\cdot C_{5}}{\theta^{\prime}}.
\]
Hence we have
\[
\limsup_{n\rightarrow+\infty}\frac{1}{N}\sum_{n=1}^{N}\mathbb{E}\left\{
f(\mathbf{q}^{n})\right\}  \leq\frac{\mathbb{E}\left\{  V(\mathbf{q}%
^{1})\right\}  }{N\theta^{\prime}}+\frac{C_{5}}{\theta^{\prime}}<+\infty
\]
which completes the proof.

\subsection{Proof of Theorem \ref{theorem:necessity}}

Considering the first case in Theorem \ref{theorem:necessity}, we define the
set of time slots in which the change of $\bar{\mu}_{i}\left(  \mathbf{q}%
(n)\right)  $ is not negligible as
\[
\mathcal{N}_{V}:=\biggl\{n:\left\Vert \boldsymbol{\bar{\mu}}\left(
\mathbf{q}^{n+1}\right)  -\boldsymbol{\bar{\mu}}\left(  \mathbf{q}^{n}\right)
\right\Vert \geq\epsilon\biggr\}
\]
for some constant $\epsilon>0$, where the superscript is again
used to denote the index of the time slot. Suppose there is some
constant $\gamma$ with $0<\gamma\leq1$ and
\[
\frac{1}{N}\sum_{n=1}^{N}\mathbb{I}\left\{  n\in\mathcal{N}_{V}\right\}
\geq\gamma+\varepsilon(N),
\]
where $\varepsilon(N)\rightarrow0$ as $N\rightarrow+\infty$. If
the difference between $\boldsymbol{\bar{\mu}}\left(
\mathbf{q}^{n+1}\right)  $ and $\boldsymbol{\bar{\mu}}\left(
\mathbf{q}^{n}\right)  $ is larger than $\epsilon$, then the
expected rate allocation
$\tilde{\mathbf{r}}_{E}^{n}:=\mathbb{E\{}\mathbf{r}^{\mathcal{P}}(\mathbf{h}^{n},\mathbf{q}^{n})\}$
and $\tilde{\mathbf{r}}_{E}^{n+1}:=\mathbb{E\{}\mathbf{r}^{\mathcal{P}%
}(\mathbf{h}^{n+1},\mathbf{q}^{n+1})\}$, which are determined by
$\bar{\boldsymbol{\mu}}\left(  \mathbf{q}^{n+1}\right)  $ and
$\bar {\boldsymbol{\mu}}\left(  \mathbf{q}^{n}\right)  $, also
have non-negligible difference. Note that this assumption is valid
only if the normal vector
$\bar{\boldsymbol{\mu}}$ is unique on every boundary point of $\mathcal{C}%
(\hat{P})$. Generally two different normal vector
$\bar{\boldsymbol{\mu}}$ and $\bar{\boldsymbol{\mu}}^{\prime}$
might lead to the same boundary point of $\mathcal{C}(\hat{P})$,
where the boundary has no unique supporting hyperplane.
Fortunately, to disprove the throughput-optimality we only need to
consider certain rate region $\mathcal{C}(\hat{P})$ whose boundary
is differentiable everywhere and the normal vector is always
unique. Since both of $\tilde{\mathbf{r}}_{E}^{n}$ and
$\tilde{\mathbf{r}}_{E}^{n+1}$ lie inside $\mathcal{C}(\hat{P})$,
for a boundary point $\mathbf{r}^{\ast}$ of
$\mathcal{C}(\hat{P})$ and the corresponding normal vector
$\boldsymbol{\mu }^{\ast}$ we have
\begin{align}
&  \boldsymbol{\mu}^{\ast
T}\tilde{\mathbf{r}}_{E}^{n}+\boldsymbol{\mu}^{\ast
T}\tilde{\mathbf{r}}_{E}^{n+1}\nonumber\\
=  &  \boldsymbol{\mu}^{\ast T}\cdot\arg\max_{\mathbf{r}'\in
\mathcal{C}(\hat{P})}\bar{\boldsymbol{\mu}}\left(
\mathbf{q}^{n}\right)
^{T}\cdot\mathbf{r}'\nonumber\\
&  +\boldsymbol{\mu}^{\ast T}\cdot\arg\max_{\mathbf{r}'\in
\mathcal{C}(\hat{P})}\bar{\boldsymbol{\mu}}\left(
\mathbf{q}^{n+1}\right)
^{T}\cdot\mathbf{r}'\nonumber\\
\leq &  2\boldsymbol{\mu}^{\ast
T}\cdot\mathbf{r}^{\ast}-\theta(\epsilon ),\nonumber
\end{align}
where $\theta(\epsilon)$ is determined by the difference $\epsilon$ between
$\bar{\boldsymbol{\mu}}\left(  \mathbf{q}^{n+1}\right)  $ and $\bar
{\boldsymbol{\mu}}\left(  \mathbf{q}^{n}\right)  $ with $\theta(\epsilon)>0$.
Considering the queue states on some even time slots $N=2,4,...$, it holds
\begin{align}
&  \boldsymbol{\mu}^{\ast T}\mathbb{E}\left\{  \mathbf{q}^{N}\right\}
\nonumber\\
=  &  \boldsymbol{\mu}^{\ast T}\sum_{n=1}^{N/2}\mathbb{E}\biggl\{\mathbf{a}%
^{n}-\mathbf{r}^{\mathcal{P}}(\mathbf{h}^{2n},\mathbf{q}^{2n})+\mathbf{a}%
^{2n+1}\biggr.\nonumber\\
&  \biggl.-\mathbf{r}^{\mathcal{P}}(\mathbf{h}^{2n+1},\mathbf{q}%
^{2n+1})\biggr\}\nonumber\\
=  &  \sum_{n=1}^{N/2}\left(  2\boldsymbol{\mu}^{\ast
T}\boldsymbol{\rho }^{\ast}-\boldsymbol{\mu}^{\ast T}\left(
\tilde{\mathbf{r}}_{E}^{2n}  +\tilde{\mathbf{r}}_{E}^{2n+1}
\right)
\mathbb{I}\left\{  2n\notin\mathcal{N}_{V}\right\}  \right. \nonumber\\
&  \left.  -\boldsymbol{\mu}^{\ast T}\left(
\tilde{\mathbf{r}}_{E}^{2n}  +\tilde{\mathbf{r}}_{E}^{2n+1}
\right)
\mathbb{I}\left\{  2n\in\mathcal{N}_{V}\right\}  \right) \nonumber\\
\geq &  \frac{N}{2}\left(  2\boldsymbol{\mu}^{\ast T}\boldsymbol{\rho}^{\ast
}-2\boldsymbol{\mu}^{\ast T}\mathbf{r}^{\ast}+\left(  \gamma+\varepsilon
(N)\right)  \theta(\epsilon)\right)  \label{ineq:necessity_C1}%
\end{align}
Suppose the expected arrival rate $\boldsymbol{\rho}^{\ast}$ is close to the
boundary point $\mathbf{r}^{\ast}$ so that
\[
\boldsymbol{\mu}^{\ast T}\mathbf{r}^{\ast}-\boldsymbol{\mu}^{\ast
T}\boldsymbol{\rho}^{\ast}<\theta^{\prime}%
\]
for some $\theta^{\prime}>0$. Combining with (\ref{ineq:necessity_C1}), it
holds
\[
\boldsymbol{\mu}^{\ast T}\mathbb{E}\left\{  \mathbf{q}(N)\right\}  >\frac
{N}{2}\left(  \gamma\theta(\epsilon)-2\theta^{\prime}+\varepsilon
(N)\theta(\epsilon)\right)
\]
Since $\varepsilon(N)\rightarrow0$ as $N\rightarrow+\infty$, if $\theta
^{\prime}<\frac{\gamma}{2}\theta(\epsilon)$, we have
\[
\lim_{N\rightarrow+\infty}\boldsymbol{\mu}^{\ast T}\mathbb{E}\left\{
\mathbf{q}(N)\right\}  =+\infty
\]
and the Markov chain is not strongly stable. Suppose the variance
$\mathbf{a}(n)$ and $\mathbf{h}(n)$ is sufficiently small, so that for some
constant $C_{A}>0$ the probability
\[
Pr\left\{  \boldsymbol{\mu}^{\ast T}\mathbf{q}(N)<C_{A}\right\}
\]
decreases sufficiently fast when $\Psi(N):=\boldsymbol{\mu}^{\ast T}%
\mathbb{E}\left\{  \mathbf{q}(N)\right\}  $ increases, i.e. $\exists
C_{B},K>0$ with
\[
K\frac{1}{\Psi(N)^{1+\Gamma}}\geq Pr\left\{  \boldsymbol{\mu}^{\ast
T}\mathbf{q}(N)<C_{A}\right\}  ,\qquad\forall\Psi(N)>C_{B},
\]
for some constant $\Gamma>0$. Define $N_{c}:=\min\{N\in\mathbb{N}%
:\Psi(N)>C_{B}\}$, the expected occupation time of the set $\mathcal{A}%
:=\left\{  \mathbf{q}:\boldsymbol{\mu}^{\ast T}\mathbf{q}<C_{A}\right\}  $ is
given by
\begin{align*}
\mathbb{E}\left\{  \eta_{\mathcal{A}}\right\}   &  =\sum_{N=1}^{\infty
}Pr\left\{  \boldsymbol{\mu}^{\ast T}\mathbf{q}(N)<C_{A}\right\} \\
&  \leq N_{c}+\sum_{N=N_{c}}^{\infty}K\left(  \frac{1}{\Psi(N)^{1+\Gamma}%
}\right) \\
&  =N_{c}+\sum_{N=N_{c}}^{\infty}K\left(  \frac{1}{\left(  \left(
\gamma\theta-\theta^{\prime}\right)  N\right)  ^{1+\Gamma}}\right) \\
&  <+\infty
\end{align*}
and the Markov chain is transient.

For the second case in Theorem \ref{theorem:necessity}, we choose the expected
arrival rate vector $\boldsymbol{\rho}^{\ast}$ close to the $j$-th corner of
the ergodic achievable rate region with
\[
r_{j}^{\ast}-\rho_{j}^{\ast}<\theta^{\prime},\qquad
\]
for the user $j\neq i$ and some constant $\theta^{\prime}>0$, where
$r_{j}^{\ast}:=\max_{\tilde{\mathbf{r}}\in\mathcal{C}(\hat{P})}\tilde{r}_{j}$.
According to (\ref{condition_necessity_C2_compl}) we have for some constant
$\epsilon>0$ and $\gamma$ with $0<\gamma\leq1$%

\[
\frac{1}{N}\sum_{n=1}^{N}\mathbb{I}\left\{  \bar{\mu}_{j}\left(
\mathbf{q}(n)\right)  \leq1-\epsilon\right\}  \geq\gamma+\varepsilon(N),
\]
where $\varepsilon(N)\rightarrow0$ as $N\rightarrow+\infty$. It implies that
for some $\theta(\epsilon)>0$, it holds
\[
\frac{1}{N}\sum_{n=1}^{N}\mathbb{I}\left\{  r_{j}^{\ast}-r_{j}^{\prime}\left(
n\right)  \geq\theta(\epsilon)\right\}  \geq\gamma+\varepsilon(N),
\]
where $r_{j}^{\prime}\left(  n\right)  :=\mathbb{E\{}\mathbf{r}^{\mathcal{P}%
}(\mathbf{h}^{n},\mathbf{q}^{n})\}$ and the corner point $[0,...,r_{j}^{\ast
},...]$ can not be achieved. Then we have
\begin{align}
&  \mathbb{E}\left\{  q_{j}^{N}\right\} \nonumber\\
=  &  \sum_{n=1}^{N}\mathbb{E}\left\{  a_{j}^{n}-r_{j}^{\mathcal{P}%
}(\mathbf{h}^{n},\mathbf{q}^{n})\right\} \nonumber\\
=  &  \sum_{n=1}^{N}\left(  \rho_{j}^{\ast}-r_{j}^{\prime}\left(  n\right)
\mathbb{I}\left\{  r_{j}^{\ast}-r_{j}^{\prime}\left(  n\right)  \geq
\theta(\epsilon)\right\}  \right. \nonumber\\
&  \left.  -r_{j}^{\prime}\left(  n\right)  \mathbb{I}\left\{  r_{j}^{\ast
}-r_{j}^{\prime}\left(  n\right)  <\theta(\epsilon)\right\}  \right)
\nonumber\\
>  &  N\left(  \gamma\theta(\epsilon)-\theta^{\prime}+\varepsilon
(N)\theta(\epsilon)\right)  .
\end{align}
Choose $\theta^{\prime}<\gamma\theta(\epsilon)$, we have
\[
\lim_{N\rightarrow+\infty}\mathbb{E}\left\{  q_{j}^{N}\right\}  =+\infty
\]
and the Markov chain is not strongly stable. Similarly we can also show that
the Markov chain is transient if the variances of $\mathbf{a}(n)$ and
$\mathbf{h}(n)$ are sufficiently small.

\bibliographystyle{IEEEbib}
\bibliography{queue_publications,own_publications,general,mimo_publications,OFDM_publications}

\end{document}